\documentclass[useAMS,usenatbib]{mn2e}
\usepackage{subfigure}
\usepackage{natbib}
\usepackage{lscape}
\usepackage[dvips]{color}
\usepackage{amssymb}
\usepackage{graphicx}
\usepackage{ulem}

\usepackage{ulem}

\title{The Properties of the Heterogeneous Shakhbazyan Groups of Galaxies in the SDSS}
\author[Capozzi D. et al. 2009]
 { Capozzi D.$^{1,2}$\thanks{E-mail:dc@astro.livjm.ac.uk}, De Filippis E.$^{1}$\thanks{betty@na.infn.it},
  Paolillo M.$^{1,4}$\thanks{paolillo@na.infn.it}, D'Abrusco R.$^{1}$\thanks{dabrusco@na.infn.it}, Longo
  G.$^{1,3,4}$\thanks{longo@na.infn.it} \\ \\
 1 - Department of Physical Sciences, University of Napoli Federico II, via Cinthia 9, 80126 Napoli, ITALY\\
 2 - Astrophysics Research Institute, Liverpool John Moores University, Twelve Quays House, Egerton Wharf,\\
  \quad \, Birkenhead, CH41 1LD, UK\\
 3 - INAF - Osservatorio Astronomico di Capodimonte, via Moiariello 16, 80131, Napoli ITALY\\
 4 - INFN - Napoli Unit, Dept. of Physical Sciences, via Cinthia 9, 80126, Napoli, ITALY}
\date{Accepted ;
      Received ;
      in original form }
\pagerange{\pageref{firstpage}--\pageref{lastpage}} \pubyear{2009}

\begin{document}

\maketitle

\label{firstpage}

\begin{abstract}
We present a systematic study of the sub-sample of Shakhbazyan groups (SHKs) covered by the 
Sloan Digital Sky Survey Data Release--5 (SDSS-5). 
SHKs probe an environment with characteristics which are intermediate between 
those of loose and very compact groups. Surprisingly, we found that several groups identifying algorithms 
({\rm e.g.} \citealp{Berlind-2006}, \citealp{Tago-2008}) miss this type of structures.
Using the SDSS-5 spectroscopic data and the photometric redshifts derived in \cite{D'Abrusco-2007}, 
we identified possible group members in photometric redshift space and derived, for each group, several 
individual properties (richness, size, mean photometric redshift, fraction of red galaxies {\rm etc.}). We also 
combined pointed and stacked {\it Rosat All Sky Survey} ({\it RASS}) data to investigate the X-ray luminosities of 
these systems.
Our study confirms that the majority of groups are physical entities with richness in the 
range 3--13 galaxies, and properties ranging between those of loose and compact groups. 
We confirm that SHK groups are richer in early-type galaxies than the surrounding
environment and the field, as expected from the morphology-density relation and from the selection of groups 
of red galaxies. 
Furthermore, our work supports the existence of two sub-classes of structures, the first one being formed
by compact and isolated groups and the second formed by extended structures.
We suggest that while the first class of objects dwells in less dense regions like the outer parts of clusters or 
the field, possibly sharing the properties of Hickson Compact Groups, the more extended structures represent a 
mixture of [core+halo] configurations and cores of rich clusters.
X-ray luminosities for SHKs are generally consistent with these results and with the expectations for the 
$L_{X}-\sigma_{v}$ relation, but also suggest the velocity dispersions reported in literature are underestimated 
for some of the richest systems.

\end{abstract}


\section{INTRODUCTION}

In spite of the fact that small groups of galaxies are the most common extragalactic environment \citep{Tully-1987},
their physical properties, origin and evolution are still poorly understood. 
This is particularly true in the low mass density regime, id est in that environmental range which
bridges the field with the richest (clusters) or most compact ({\rm e.g.} Hickson Compact groups, hereafter HCGs) 
structures, mainly due to observational selection effects which render the poor and less compact 
structures much more difficult to detect and study, especially at intermediate and high 
redshift.

The lack of statistically complete and well measured samples of groups at different redshifts 
is very unfortunate since they are needed to constrain all models for the formation and evolution 
of cosmic structures.
For instance, in the hierarchical models it would be crucial to measure the epoch of groups formation
and to discriminate whether it does or doesn't exist a down-sizing effect for which high-mass groups 
at high z form before lower-mass ones at lower z.
Other crucial issues are related to the dynamical status of groups, to their relaxation time and to how 
the evolution of the structure affects that of their galaxy members. 

Groups of galaxies are very numerous and 
show spatial densities of about $10^{-3}\div 10^{-5}\ {\rm h^{3}\ groups\ Mpc^{-3}}$ \citep{Bahcall-1999}. Their gravitational 
potential wells are about as deep as that of individual galaxies, whose random velocities in such systems are a few hundreds
of ${\rm km\ s^{-1}}$. Under these conditions, galaxies strongly interact both among themselves and with the global 
potential of the group. Therefore, groups of galaxies are collisional systems evolving toward virial equilibrium 
through collisions and collisionless interactions among their member galaxies.

Groups of galaxies seem to evolve assuming different configurations: loose, [core+halo] and compact. Compact 
configurations form in the last phases of the evolving process and their final byproduct is a giant galaxy 
surrounded by a hot X-ray emitting gas halo, with properties similar to those of a giant elliptical galaxy, known 
as Fossil Group \citep{Vikhlinin-1999, Jones-2003, deOliveira-2006, deOliveira-2007}.
Numerical simulations have extensively shown that structures of high spatial densities undergo 
fast dynamical evolution and merging \citep{Barnes-1985,Barnes-1989,Mamon-1987,Bode-1993,Diaferio-1993,Diaferio-1994,
Governato-1996}. The expected number of groups inferred through theoretical 
estimates of crossing times is in contrast with the observed one. The Secondary Infall Scenario suggested by \citep{Gunn-1972} 
and then reconsidered by \citep{Mamon-1996, Mamon-2000, Mamon-2007}, allows a secondary collapse of
galaxies surrounding the formed structures. This secondary aggregation provides the structures a way to increase 
their lifetimes, according with the observed number of groups of galaxies \citep{Diaferio-1994}.\\

In this paper we adopt a homogeneous and systematic approach to the study of Shakhbazyan groups of galaxies 
(hereafter SHK), using the unique SDSS data-set, in order to  analyse their properties and compare them to those 
of the environment and of the well studied HCGs. 

Shakhbazyan groups of galaxies were originally defined as compact groups of mainly red compact galaxies and 
selected by visually inspecting the printed version of the First Palomar Sky Survey (POSS) using rather empirical 
and ill defined selection criteria:

\noindent - They must contain 5-15 member galaxies.

\noindent - Each galaxy's apparent magnitude in the POSS red band must be comprised 
between $14^{m}\div 19^{m}$.

\noindent - They are compact, id est the relative distances of the member galaxies 
are typically only $3\div 5$ times the characteristic diameter of a member galaxy.

\noindent - Almost all galaxies must be extremely red; there must not be more than $1-2$ blue galaxies.

\noindent - Galaxies are compact (high surface brightness and border not diffuse).

\noindent - The group must be isolated.

The search lead to a long series of papers \citep{Shakhbazyan-1973, Shakhbazyan-1974,Baier-1974,Petrosian-1974, Baier-1975,
Baier-1976, Bayer-1976, Baier-1978, Baier-1979, Petrosian-1978} identifying a total of 377 groups which, due to the poor 
resolution of the POSS and to the compactness requirement, appeared initially to be strongly contaminated by stars mistaken 
for galaxies and, furthermore, resulted affected by many systematics \citep{Thompson-1976, Kirshner-1980, Kodaira-1988, delOlmo-1995}. 
In addition, among the selection criteria, there is no explicit criterion on the magnitude range of the member 
galaxies, which hence depends on the magnitude of the group's brightest galaxy.
This suggests that SHK groups may be a heterogeneous collection of loose nearby groups and denser distant ones. 
For these reasons, until a few years ago, the SHK sample has not received as much attention as 
other more homogeneous and better defined samples such as, for instance, the Hickson's compact groups one. 

Detailed photometry by \cite{Thompson-1976, Kodaira-1988, Kodaira-1991, delOlmo-1995} showed that in most cases, those which were 
believed compact and very red galaxies ({\rm e.g.} \citealp{Robinson-1973, Mirzoyan-1975, Tiersch-1976}), were rather normal elliptical and S0 galaxies with slightly redder 
colours ($\Delta ({\it V-R})\sim 0.2$) than field ones. 
Furthermore, even though some contamination by stars is indeed present, many of the objects initially suspected 
to be stars were found to be galaxies. 
The extensive and detailed studies by Tiersch,   
\citep{Tiersch-1976,Tiersch-1993, Tiersch-1994, Tiersch-1995,Tiersch-1996a, Tiersch-1996b, Tiersch-1999a, Tiersch-1999b,Tiersch-2002}, 
Tovmassian \citep{Tovmassian-1998, Tovmassian-1999a, Tovmassian-1999b, Tovmassian-2001a, Tovmassian-2003a, Tovmassian-2003b, Tovmassian-2004, Tovmassian-2005a, Tovmassian-2005b, Tovmassian-2005c, Tovmassian-2006, Tovmassian-2007} 
and collaborators of a sub-set of 44 
groups, have shown that SHKs form a rather intriguing class of physically bound and moderately compact structures. The 
spatial densities of SHKs found in these works, span a wide range: from slightly higher than those of loose groups 
($10-10^2\ {\rm gal\ Mpc^{-3}}$) up to values comparable to the cores of rich clusters or HCGs 
($10^4-10^5\ {\rm gal\ Mpc^{-3}}$) for $m\lesssim 19\ {\rm mag}$. 
Therefore the observed spatial densities of SHKs imply that they may be at different stages of dynamical and morphological 
evolution and can be used both to probe the effects of environment on galaxy evolution and to constrain the formation 
mechanisms of low richness structures.\\ 
The paper is structured as it follows. In Sect.~\ref{sec:thedata} we describe the data. 
In Sect.~\ref{sec:themethod} and \ref{sec:groups-properties} we describe in some detail the method used to derive the 
observable quantities, while in Sect.~\ref{sec:individual} we summarize the main properties of each group in our 
sample individually. Sect.~\ref{sec:Results} is dedicated to the discussion of the global properties of the sample,
while in Sect.~\ref{sec:conclusions} we draw our conclusions.

Throughout this paper we assume a standard cosmology with $H_{0}= 70\ {\rm km\ s^{-1}\ Mpc^{-1}}$, $\Omega_{m}=0.3$ and 
$\Omega_{\Lambda}=0.7$.

\section{THE DATA}
\label{sec:thedata}
For our analysis we made use of the Sloan Digital Sky Survey - Data release 5 
(hereafter SDSS-DR5) public archive.
The SDSS covers 8,000 sq. deg. of the celestial 
sphere in 5 bands (with effective wavelengths, {\it u}=3540 {\AA}, {\it g}=4760 {\AA}, 
{\it r}=6280 {\AA}, {\it i}=7690 {\AA}, {\it z}=9250 {\AA}) and is complemented by an extensive 
spectroscopic survey providing spectroscopic redshifts for $674,749$ 
galaxies\footnote{A detailed description of the survey can be found at http://www.sdss.org/dr5.}.

The spectroscopic survey is almost complete for galaxies with $r<17.77$, while 
at fainter light levels it includes mainly Luminous Red Galaxies (LRG) \citep{Eisenstein-2001}.
214 SHKs lie in the sky area covered by the SDSS-DR5. 
Among these 214 groups, we selected those for which a first order estimate of the distance was possible, averaging 
either spectroscopic redshifts from literature, or spectroscopic redshifts available in the SDSS-DR5, for 
all galaxies contained in the catalogue by \citep{Stoll-1993a, Stoll-1993b, Stoll-1994a, Stoll-1994b, Stoll-1996a, 
Stoll-1996b, Stoll-1996c, Stoll-1997a, Stoll-1997b, Stoll-1997c}\footnote{Since the submission of the paper SDSS-DR7 was released. Nevertheless we note that DR7 has a different footprint from DR5 and thus would not increase the number of available redshifts per group. However it may be used in future works to extend the sample size.}. The same process has been used to derive the 
centroids of the groups using the coordinates of the single galaxies in the catalogue by Stoll et al. \\ 
This selection resulted in a spectroscopic sub-sample of 58 SHK groups on which this paper is focused. Groups' 
positions and mean spectroscopic redshift estimates are listed in Table~\ref{tab:Table1}.

\footnotesize
\begin{table*}
\begin{center}
\begin{tabular}{lccrllcl}
\hline
\hline
 {\bf SHK} & {\bf RA} & {\bf DEC} & {$\mathbf{\overline{z}_{{\rm {\bf spec}}}}$} & {\bf n} &  
 {$\mathbf{\overline{z}_{{\rm {\bf phot}}}}$}& {$\mathbf{\epsilon(\overline{z}_{{\rm {\bf phot}}})}$}& 
 {\bf References} \\
\hline
1      &     10:55:05.70  &   +40:27:30.0  &   0.117  &    7  &   0.10    &   0.01	 &  Kirshner, R. P. \& Malumuth, E. M., 1980, ApJ, 236, 366		   \\ 
5      &     11:17:06.75  &   +54:55:10.3  &   0.139  &    5  &   0.14    &   0.01	 &  Stoll, D., et al., 1993-1999, AN					   \\ 
6      &     11:18:49.10  &   +51:44:37.2  &   0.079  &    3  &   0.09    &   0.02	 &  SDSS DR5								   \\ 
8      &     16:03:41.02  &   +52:21:13.7  &   0.110  &    6  &   0.12    &   0.02	 &  Tovmassian, M., et al., 2005, A\&A, 439, 973			   \\ 
10     &     14:10:48.93  &   +46:15:54.2  &   0.130  &    5  &   0.13    &   0.01	 &  SDSS DR5								   \\ 
11     &     14:11:06.15  &   +44:43:05.8  &   0.095  &    4  &   0.10    &   0.01	 &  SDSS DR5								   \\ 
14     &     14:25:19.75  &   +47:15:09.5  &   0.073  &    7  &   0.08    &   0.01	 &  Tovmassian, M., et al., 2005, A\&A, 439, 973			   \\ 
19     &     13:28:30.20  &   +15:50:25.6  &   0.069  &    4  &   0.02    &   0.02	 &  Tovmassian, M., et al., 2005, A\&A, 439, 973			   \\ 
22     &     15:45:43.74  &   +55:06:58.8  &   0.082  &    4  &   0.06    &   0.02	 &  Tovmassian, M., et al., 2005, A\&A, 439, 973			   \\ 
29     &     16:08:42.16  &   +52:26:19.0  &   0.035  &    1  &   0.17    &   0.02	 &  Tovmassian, H. M., et al., 1999, ApJ, 523, 87			   \\ 
31     &     00:58:17.99  &   +13:54:38.7  &   0.187  &    5  &   0.18    &   0.02	 &  Tovmassian, H. M., et al., 2003, RMxAA, 39, 275  			   \\ 
54     &     10:40:32.96  &   +40:14:38.3  &   0.086  &    6  &   0.06    &   0.01	 &  SDSS DR5								   \\ 
55     &     10:43:34.89  &   +48:22:30.4  &   0.143  &    1  &   0.13    &   0.01	 &  SDSS DR5								   \\ 
57     &     10:45:26.74  &   +49:31:37.9  &   0.174  &    4  &   0.17    &   0.02	 &  SDSS DR5								   \\ 
60     &     11:24:35.75  &   +40:25:43.0  &   0.108  &    4  &   0.16    &   0.01	 &  SDSS DR5								   \\ 
63     &     11:29:34.18  &   +42:26:15.8  &   0.181  &    1  &   0.21    &   0.01	 &  SDSS DR5								   \\ 
65     &     11:30:48.45  &   +35:02:49.4  &   0.185  &    2  &   0.14    &   0.02	 &  SDSS DR5								   \\ 
70     &     12:01:18.93  &   +41:14:19.0  &   0.109  &    2  &   0.20    &   0.01	 &  SDSS DR5								   \\ 
74     &     14:21:06.05  &   +43:03:46.7  &   0.104  &    8  &   0.12    &   0.02	 &  Tovmassian, H. M., et al., 2005, RMxAA, 39, 275			   \\ 
95     &     08:28:36.12  &   +50:17:53.2  &   0.079  &    2  &   0.07    &   0.01	 &  SDSS DR5								   \\ 
96     &     08:37:54.46  &   +52:37:23.6  &   0.097  &    1  &   0.11    &   0.01	 &  SDSS DR5								   \\ 
104    &     09:27:13.60  &   +52:58:40.5  &   0.167  &    5  &   0.11    &   0.01	 &  Tovmassian, H. M., et al., 2007, RMxAA, 43, 45			   \\ 
120    &     11:04:28.47  &   +35:52:50.6  &   0.070  &    7  &   0.13    &   0.02	 &  Tovmassian, H. M., et al., 2007, RMxAA, 43, 45			   \\ 
123    &     11:44:48.31  &   +57:31:52.5  &   0.117  &    5  &   0.09    &   0.02	 &  SDSS DR5								   \\ 
128    &     13:19:55.36  &   +55:45:21.7  &   0.145  &    2  &   0.14    &   0.02	 &  SDSS DR5								   \\ 
152    &     09:39:05.97  &   +01:56:46.4  &   0.134  &    2  &   0.48    &    --	 &  SDSS DR5								   \\ 
154    &     11:22:53.28  &   +01:06:46.3  &   0.073  &    6  &   0.08    &   0.02	 &  Tiersch, H. et al., 2002, A\&A, 392, 33				   \\ 
181    &     08:28:01.06  &   +28:15:56.3  &   0.093  &    8  &   0.10    &   0.02	 &  Tovmassian, H. M., et al., 2004, A\&A, 415, 803			   \\ 
184    &     09:08:07.50  &   +30:36:33.4  &   0.154  &    2  &   0.05    &    --	 &  SDSS DR5								   \\ 
186    &     09:22:52.08  &   +28:55:25.2  &   0.077  &    3  &   0.08    &   0.02	 &  SDSS DR5								   \\ 
188    &     09:56:59.23  &   +26:10:27.3  &   0.080  &    7  &   0.08    &   0.02	 &  Tovmassian, H. M., et al., 2005, RMxAA, 39, 275			   \\ 
191    &     10:48:09.20  &   +31:28:51.7  &   0.118  &    12 &   0.13    &   0.01	 &  Tovmassian, H. M., et al., 2005, AN, 326, 362			   \\ 
202    &     12:19:47.53  &   +28:24:13.5  &   0.028  &    9  &   0.02    &   0.01	 &  Stoll, D., et al., 1993-1999, AN					   \\ 
205    &     12:35:23.55  &   +27:34:45.7  &   0.096  &    7  &   0.12    &   0.02	 &  Stoll, D., et al., 1993-1999, AN					   \\ 
213    &     13:45:12.24  &   +26:53:44.0  &   0.058  &    1  &   0.10    &   0.02	 &  Stoll, D., et al., 1993-1999, AN					   \\ 
218    &     14:33:39.13  &   +26:41:02.6  &   0.095  &    1  &   0.09    &   0.02	 &  Tovmassian, H. M., et al., 1999, ApJ, 523, 87			   \\ 
223    &     15:49:42.86  &   +29:09:37.5  &   0.083  &    10 &   0.09    &   0.02	 &  Tovmassian, H. M., et al., 2007, RMxAA, 43, 45			   \\ 
229    &     09:00:43.69  &   +33:45:01.4  &   0.124  &    1  &   0.03    &   0.01	 &  SDSS DR5								   \\ 
231    &     10:01:41.81  &   +38:18:44.1  &   0.146  &    1  &   0.14    &   0.01	 &  SDSS DR5								   \\ 
237    &     11:05:29.36  &   +38:00:48.6  &   0.030  &    1  &   0.08    &   0.02	 &  Stoll, D., et al., 1993-1999, AN					   \\ 
245    &     12:24:45.80  &   +31:57:17.3  &   0.063  &    5  &   0.05    &   0.02	 &  Kodaira, K., et al., 1991, PASJ, 43, 169				   \\ 
248    &     13:12:16.40  &   +36:11:17.4  &   0.271  &    1  &   0.18    &   0.01	 &  Tovmassian, H. M., et al., 1999, ApJ, 523, 87			   \\ 
251    &     13:36:54.80  &   +36:49:37.7  &   0.061  &    6  &   0.05    &   0.02	 &  Tovmassian, H. M., et al., 2005, RMxAA, 39, 275					   \\ 
253    &     13:52:23.70  &   +37:30:59.7  &   0.073  &    1  &   0.08    &   0.02	 &  Stoll, D., et al., 1993-1999, AN					   \\ 
254    &     13:56:24.79  &   +35:11:10.4  &   0.170  &    3  &   0.16    &   0.01	 &  SDSS DR5								   \\ 
258    &     15:23:39.93  &   +32:24:12.8  &   0.032  &    1  &   0.17    &   0.02	 &  SDSS DR5								   \\ 
344    &     08:47:32.54  &   +03:42:01.0  &   0.077  &    5  &   0.07    &   0.01	 &  Tovmassian, H. M., et al., 2004, A\&A, 415, 803			   \\ 
346    &     09:15:10.16  &   +05:14:21.4  &   0.135  &    1  &   0.13    &   0.02	 &  Tovmassian, H. M., et al., 1999, ApJ, 523, 87			   \\ 
348    &     09:26:35.17  &   +03:26:39.7  &   0.088  &    8  &   0.09    &   0.02	 &  Tovmassian, H. M., et al., 2005, RMxAA, 39, 275					   \\ 
351    &     11:10:19.20  &   +04:47:31.8  &   0.030  &    4  &   0.04    &   0.02	 &  SDSS DR5								   \\ 
352    &     11:21:37.95  &   +02:53:20.2  &   0.049  &    1  &   0.05    &   0.02	 &  Tovmassian, H. M., et al., 1999, ApJ, 523, 87			   \\ 
355    &     13:12:11.33  &   +07:18:28.8  &   0.093  &    4  &   0.10    &   0.01	 &  Stoll, D., et al., 1993-1999, AN					   \\ 
357    &     13:42:10.29  &   +02:13:42.5  &   0.077  &    21 &   0.08    &   0.02	 &  SDSS DR5								   \\ 
358    &     14:23:46.29  &   +06:35:05.4  &   0.050  &    3  &   0.06    &   0.02	 &  SDSS DR5								   \\ 
359    &     14:29:56.51  &   +18:50:20.0  &   0.033  &    1  &   0.10    &   0.02	 &  Tovmassian, H. M., et al., 1999, ApJ, 523, 87			   \\ 
360    &     15:41:26.72  &   +04:44:09.7  &   0.108  &    8  &   0.12    &   0.01	 &  Tiersch, H. et al., 2002, A\&A, 392, 33				   \\ 
371    &     11:43:33.32  &   +21:53:57.0  &   0.130  &    1  &   0.13    &   0.02	 &  Tovmassian, H. M., et al., 1999, ApJ, 523, 87			   \\ 
376    &     13:56:34.42  &   +23:21:48.5  &   0.067  &    10 &   0.02    &   0.02	 &  Tovmassian, H. M., et al., 2003, A\&A, 401, 463			   \\ 
\hline									  						        						      
\end{tabular}								  						        						      
\caption{{\bf SHKs in the literature.}
Column 1: identification number; column 2 and 3: right ascension and declination (J2000); column 4:    						      
group's mean spectroscopic redshift from literature (see col.8); column 5: number of available
accordant spectroscopic redshifts (SDSS and literature) per group for galaxies in the SHK catalogue 
by Stoll et al.; column 6 and 7: group's mean photometric redshift and its error; column 8: spectroscopic 
redshift reference.}													        						      
\label{tab:Table1} 							    
\end{center}
\end{table*}
\normalsize

For each of these groups we extracted from DR5 all objects within a 
projected distance of 3 {\rm Mpc} 
from the centroid of the group, in order to be able to study the prospective over-density in relation 
with the environment in which it dwells. This approach is different from the one often used to 
study compact groups such as HCGs, whose selection criteria \citep{Hickson-1982} completely isolates them 
from their environment, causing a loss of information about possible extended structures in which they 
are embedded \citep{Palumbo-1995,deCarvalho-2000}.

Due to the compactness selection criterion used, the original member lists of the Shakhbazyan groups 
were contaminated by stars mistaken for compact galaxies. 
The better resolution of the SDSS with respect to the older POSS data should itself 
ensure that many misclassified objects are removed from our lists. As an additional
check we compared the reliability of the Star/Galaxy classification provided by 
the SDSS classification algorithm (PHOTO) \citep{Lupton-2001} with a second indicator introduced by 
\cite{Yasuda-2001}, relying on a series of SDSS photometric parameters and flags. 
The comparison lead to a rate of misclassification $<1$ per cent which does not affect our results. Objects with 
conflicting classifications were excluded from our final catalogue.

Photometric redshifts for all galaxies within 3 {\rm Mpc} were then extracted from the 
catalogue by \cite{D'Abrusco-2007}, which contains photometric estimates for
objects in the SDSS brighter than $r=21.0$, with $z<0.55$, and with an accuracy of 
$\epsilon(z_{{\rm phot}})=0.02$~\footnote{The photometric redshifts catalogues are publicly available at:
http://people.na.infn.it/$\sim$astroneural/catalogues.html.}.
We refer to the original paper for further details on how these redshifts were 
evaluated and for a thorough discussion of their accuracy.
The decision of using photometric redshifts from \citet{D'Abrusco-2007} has been driven by three main reasons:
1) their accuracy ($\epsilon(z_{{\rm phot}})=0.02$) is comparable or better than the SDSS estimates Photoz 
(\citet{Csabai-2003}, $\epsilon(z_{{\rm phot}})=0.035$) and Photoz2 (\citet{Oyaizu-2008}, $\epsilon(z_{{\rm phot}})=0.021$); 
2) the low redshift range of our sample of SHK groups is accurately covered by the \citet{D'Abrusco-2007} redshifts.
PhotoZ and PhotoZ2 are instead less complete at low redshifts, due to the strict selection criteria of their 
samples.
3) the uncertainties in the D'Abrusco photo-z's show a nearly gaussian distribution, with very few catastrophic 
outliers. This makes the choice of the latter sample preferable for our application.

\begin{figure*}
\includegraphics[height=0.355\textheight]{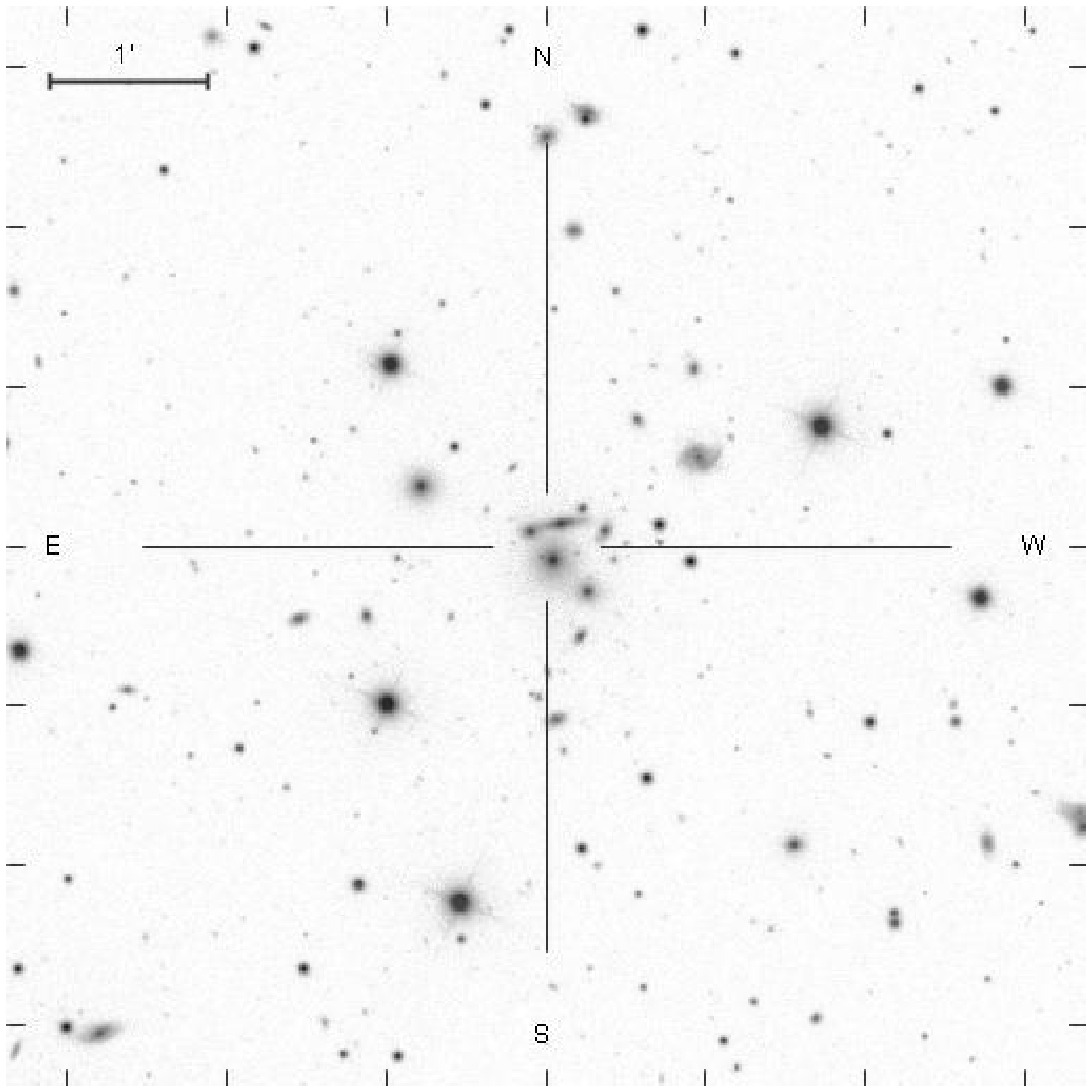}
\hfill
\includegraphics[height=0.365\textheight]{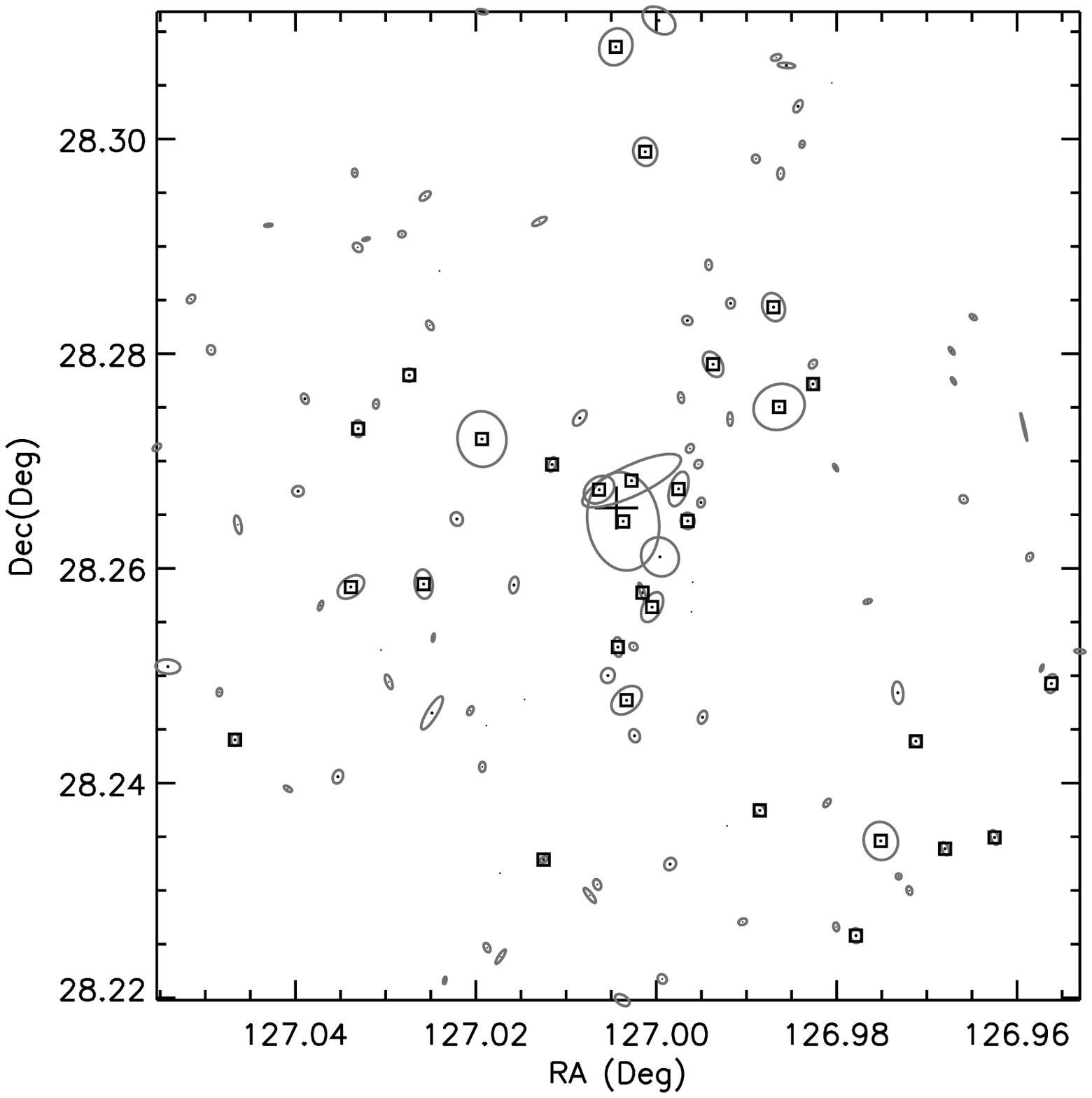}
\caption{ Diagnostic plots for the group SHK 181. 
Left-hand panel: group optical SDSS-DR5 image in the{\it r} band.
Right-hand panel: map of the projected galaxy distribution within the central $300\ {\rm kpc}$. Squares 
mark objects whose photometric redshifts lie within $3\epsilon(z_{{\rm phot}})$ of the group mean spectroscopic 
redshift.} 
\label{fig:Fig1}
\end{figure*}

\section{THE METHOD}
\label{sec:themethod}
In order to address and investigate the properties of our sample of groups, we first checked each group 
individually to verify the galaxy membership and compare it with the results presented in the literature.
We then derived a number of diagnostic tools (projected number density maps, radial profiles, 
redshift distributions, colour--magnitude diagrams) as described in more detail in the following sections. 
We labelled a group as a real structure if it satisfied the following two criteria:\\
- presence of an excess of galaxies in the projected distribution and radial profile plots;\\
- presence of a peak in the spectroscopic/photometric redshift distribution;\\
Since early-type galaxies in clusters and often in groups sit on well defined sequences (hereafter Red Sequence, 
RS) in colour--magnitude diagrams \citep[cf.][and references therein]{Tanaka-2005,Bernardi-2003}, as an additional 
check we verified the presence of a well-defined colour--magnitude relation (hereafter CMR). This check is 
intended only to provide additional support to the presence of a group, as this relation is not necessarily well 
defined in groups of galaxies but depends on their early-type galaxy content and on their evolutionary status.\\
It must be stressed that the presence of a peak in projected spatial or redshift distribution doesn't 
prove its physical reality, since, as showed by \citet{Mamon-1986,Walke-1989,McConnachie-2008a} and 
\citet{Diaz-2008}, they might be chance alignments of galaxies along the line of sight within looser groups.
With ``real groups'' we hence indicate those structures which are not obvious superpositions
of background and foreground galaxies. 

To investigate the global properties of the sample, we further estimated for each group, their richness, 
early-type fraction and angular size. Results for all groups are summarised in Table \ref{tab:Table2} 
\footnote{Diagnostic diagrams for all studied SHK groups can be found at: http//www.astro.ljmu.ac.uk/$\sim$dc/Groups/SHK\_Groups.}.
All the properties, unless explicitly stated, are estimated considering only the galaxies satisfying the
following conditions:

\begin{itemize}
\item[1)] Radial distance selection: galaxies within a fixed radial distance 
(either $R<150\ {\rm kpc}$ or $R<500\ {\rm kpc}$) in the group rest-frame;
\item[2)] Magnitude selection criterion: galaxies such that $(r-r_{1})^{0}<3$ where $r^{0}_{1}$ is the 
magnitude of the brightest galaxy within $150\ {\rm kpc}$;
\item[3)] Photometric redshift criterion: galaxies whose photometric redshift satisfy the condition 
$|z_{{\rm phot}}-\overline{z}_{{\rm spec}}|\leq 3\epsilon(z_{{\rm phot}})$;
\end{itemize}
Since $\epsilon(z_{{\rm phot}})=0.02$, last condition amounts to an uncertainty of $\sim
18000\ {\rm km\ s^{-1}}$, i.e. $\sim 250\ {\rm Mpc}$. Therefore we expect significant contamination by
interlopers, which are later removed through a statistical approach. Notwithstanding this, the latter condition 
is very effective in increasing the S/N ratio, removing a large fraction of foreground and background galaxies.  

Finally, to compare SHKs with the well studied HCGs, in an homogeneous way, we derived all the quantities 
discussed below also for a sub-sample of 15 HCGs (HCG 17, 35, 43, 45, 47, 49, 50, 52, 57, 60, 66, 70, 71, 72, 
73, 74, 75, 76, 81, 82) with $\overline{z}_{{\rm spec}}\geq 0.03$ falling within the SDSS area. 
The lower limit for the redshift is applied in order to have overlapping redshift ranges with our SHK sample, 
and to limit the area under investigation which, if too large, would reduce too much the S/N ratio.\\
Based on the analysis of the radial distributions, we defined three regions around each group's centroid: 
i) an inner one, within a projected radius of $150\ {\rm kpc}$; 
ii) an annular, intermediate region, within a radius $150\ {\rm kpc}<r<1\ {\rm Mpc}$; 
iii) an annular outer region comprised between $2$ and $3\ {\rm Mpc}$ which 
defines what we shall call the ``local background''.

We point out that a proper characterization of the sample properties should be performed using a physical 
aperture such as the virial radius (for its derivation see e.g appendix A of \citealp{Mauduit-2007}). 
Unfortunately at the moment the spectroscopic data available for SHK groups are very sparse and 
not sufficient to derive these apertures.\footnote{We are currently conducting a spectroscopic campaign to collect 
a proper sample of spectroscopic redshift and refine our analysis.}

\begin{figure}
\includegraphics[height=0.365\textheight]{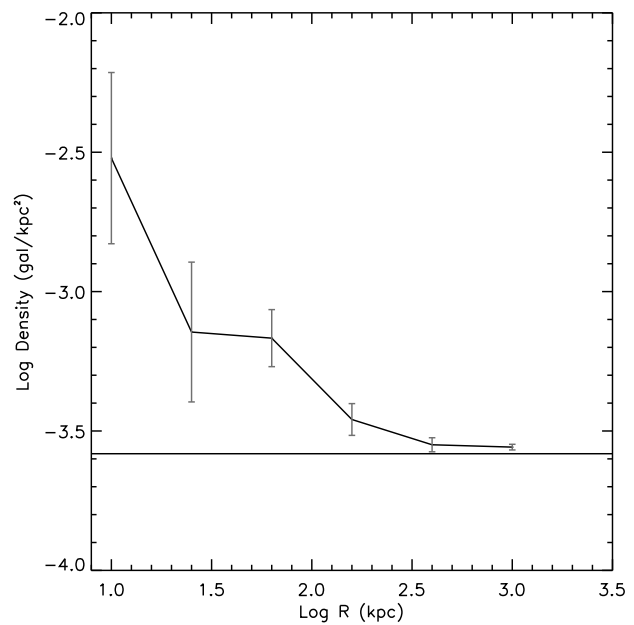}
\caption{Logarithmic radial surface number density profile for group SHK 181. The full black line represents the 
background level.}
\label{fig:Fig2}
\end{figure}

\subsection{2-D Observables} 
\label{subsec:2D_diagnostics} 
In order to test the presence of peaks in the galaxy distributions, we study their spatial distribution within 
the central 300 {\rm kpc}. Candidate member objects are classified according to their photometric redshifts, in order to 
highlight the three-dimensional distribution of the candidate group members. We identify as candidate members all 
galaxies within $3\epsilon(z_{{\rm phot}})$ from the mean spectroscopic redshift. In the upper panels
of Fig.(\ref{fig:Fig1}) an example for SHK~181 (right panel), together with the corresponding SDSS image (left panel), 
are shown. For each group we also derived a surface density map of all galaxies 
within a region of $1\ {\rm Mpc}$ radius, with a resolution of $75\ {\rm kpc}$ in the group rest-frame. 
Density maps allow to validate the position of the nominal centroid of the group and to check for the presence of 
possible sub-clumps or more extended structures.
A further tool, to verify the presence of a galaxy over-density is the number density radial profile.
We measured radial profiles, for all SHK groups out to $3\ {\rm Mpc}$ 
(Fig.~\ref{fig:Fig2}), in order to sample the local background even in the presence of extended haloes. 

We stress that even when the observables are projected quantities, we made use of the photometric redshift 
information, to remove foreground/background contaminants, thus effectively increasing the S/N ratio of our 
observables.

\begin{figure*}
\begin{center}
\includegraphics[height=0.36\textheight,clip]{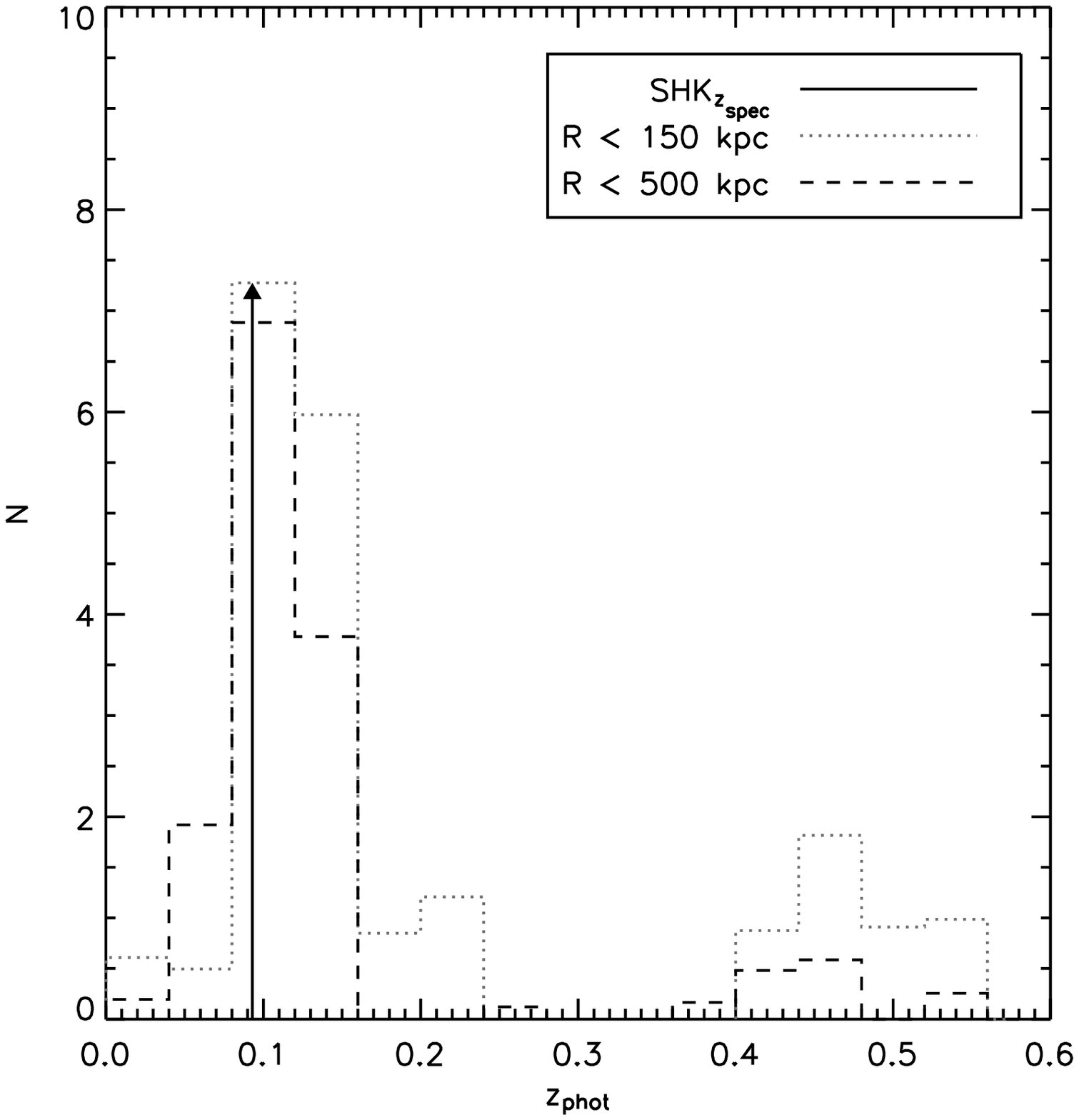}
\hfill
\includegraphics[height=0.36\textheight,clip]{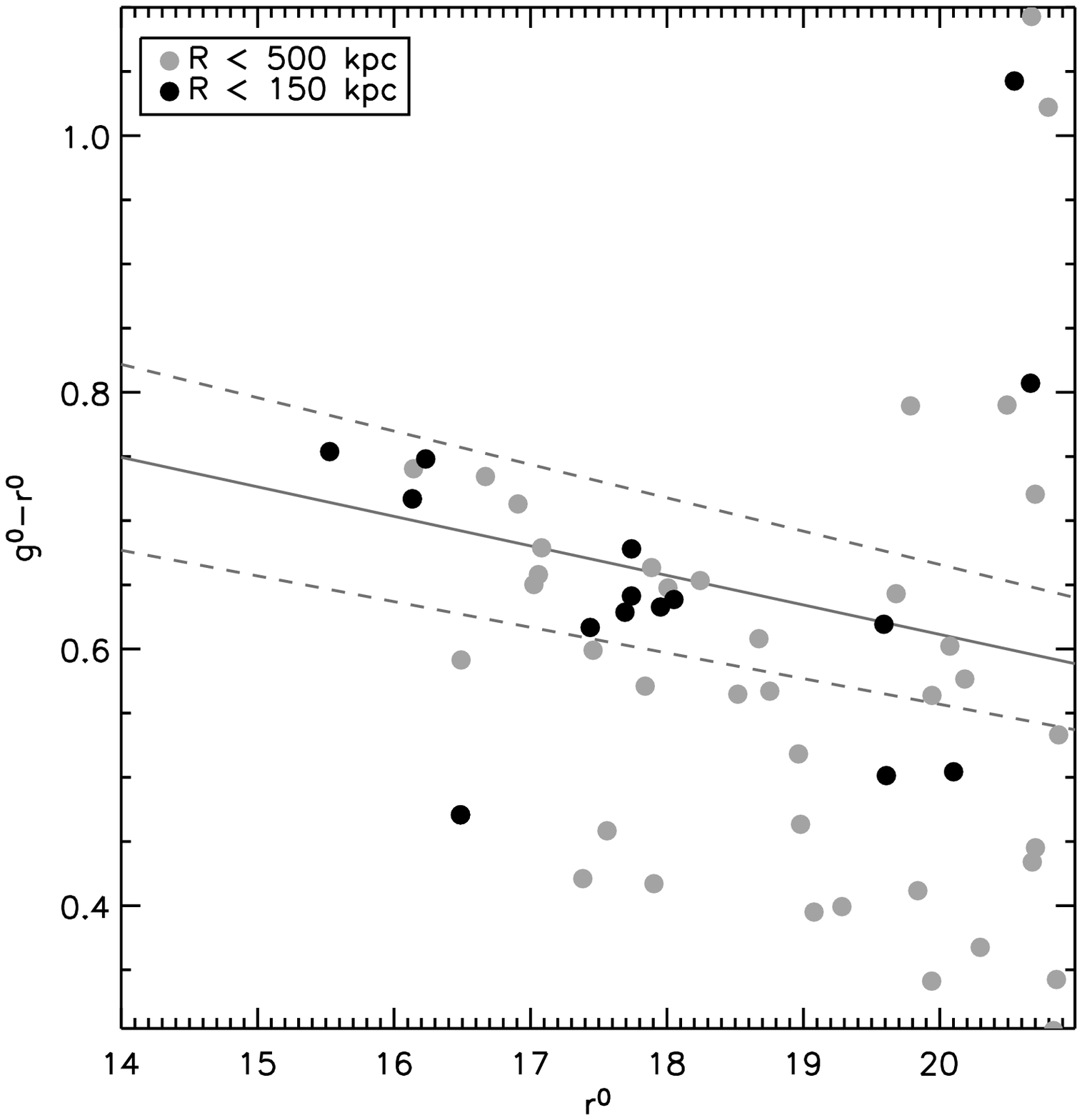}
\caption{Diagnostic plots for the group SHK 181. 
Left panel: background subtracted photometric redshift distributions for galaxies with 
$R<150\ {\rm kpc}$ (dotted line) and with $R<500\ {\rm kpc}$ (dashed line). The solid vertical line represents the 
group's mean spectroscopic redshift taken from literature.
Right panel: colour--magnitude diagram for all the galaxies contained within $R=500\ {\rm kpc}$ from the SHK 181 
group's centroid and satisfying the criteria 2 and 3 defined in \S (\ref{sec:themethod}). Galaxies within the 
inner region are represented by black dots. The solid line represents the colour--magnitude relation obtained by 
Bernardi et al. (2003) at the group's redshift, while the dashed ones indicate the scatter in its slope.}
\label{fig:Fig3}
\end{center}
\end{figure*}

\subsection{Observables in Photometric Redshift Space}
\begin{figure}
\begin{center}
\includegraphics[width=0.4\textwidth,angle=0,clip]{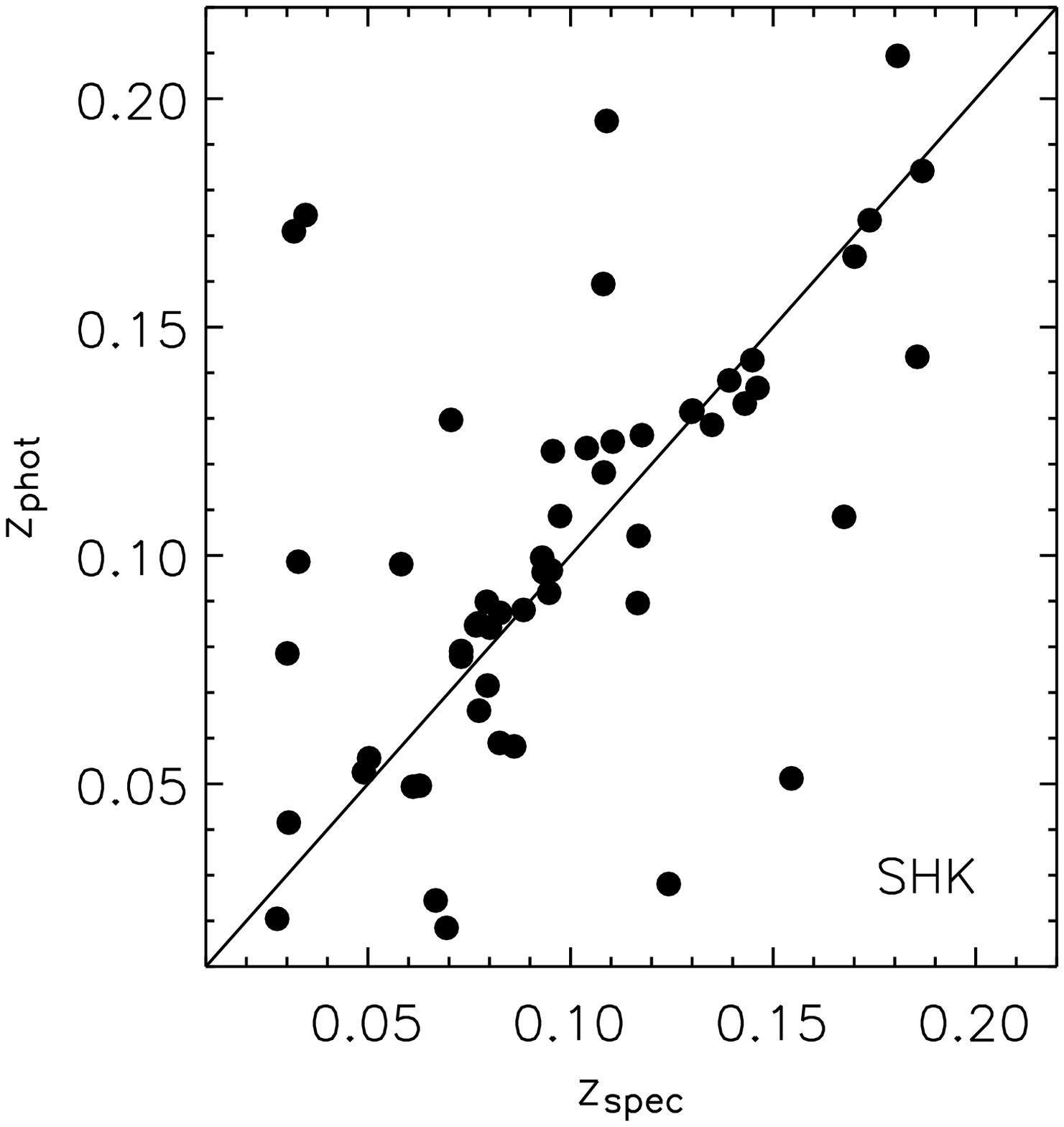} 
\hfill
\includegraphics[width=0.4\textwidth,angle=0,clip]{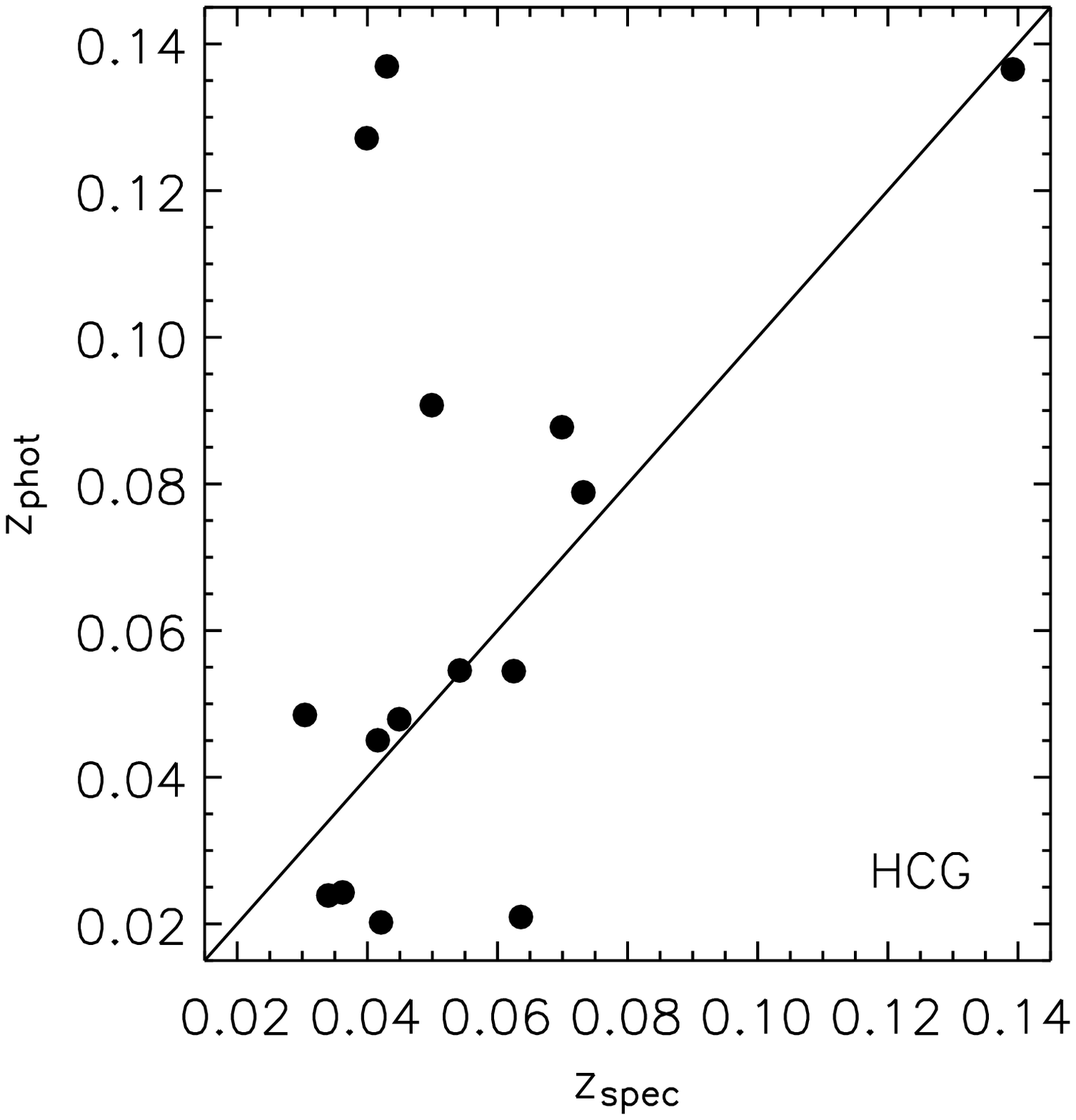} 
\caption{Comparison between the spectroscopic and photometric redshift estimates for the SHKs (upper panel) and 
HCGs (lower panel) samples. Note that in the most extreme cases, the disagreement between the two estimates 
is mainly due to the poor spectroscopic coverage or superposition of structures along the line of sight, and not 
to the performance of the photometric technique.}
\label{fig:Fig4}
\end{center}
\end{figure}

Projected data are not sufficient to identify structures of low multiplicity, due to the contamination
introduced by foreground/background galaxies. Unfortunately, because of fiber crowding, SDSS spectroscopic 
redshifts are too sparsely distributed to cover a significant number of objects in each group.
Photometric redshifts, therefore, provide a unique tool to investigate the group redshift distributions. 
Fig.~(\ref{fig:Fig3}, left panel) shows an example of the excess in the photometric redshift distribution, 
revealing the presence of a 3D structure at the group position. We then estimated the mean photometric redshift of 
each group as the average $\overline{z}_{{\rm phot}}$ of all galaxies within $\Delta z=2\epsilon(z_{{\rm phot}})=0.04$ from the 
main peak of the distribution, within a projected distance of 150 {\rm kpc}. A comparison between the mean spectroscopic 
and photometric redshift estimates for the SHKs and HCGs is shown in Fig. \ref{fig:Fig4}.
It needs to be stressed that for all those groups which have not been subjected to targeted observations, the mean 
spectroscopic redshift is often based just on a few, or worse  on only one galaxy, which may not be group member.
This partly explains the discrepancies between the mean spectroscopic and photometric redshifts listed in Table 
\ref{tab:Table1} and discussed in more detail in \S (\ref{sec:individual}). 
Note that our photometric redshift selection criterion tends to sample larger volumes at higher redshifts, thus 
over-weighting background galaxies and skewing the overall $z_{{\rm phot}}$ distribution (including groups with 
compatible redshifts) toward slightly higher values. This effect, that depends on the redshift value and 
uncertainty, results in an average correction of $\Delta z\simeq 0.01$ which is already included in 
Figure \ref{fig:Fig4}. 

In the end $77$ per cent of the groups have consistent spectroscopic and
photometric redshifts. However $23$ per cent of them have discordant values, either due to the small number of
spectroscopic data or to the presence of different structures superimposed along the line of sight.
Since an incorrect estimate of the group mean redshift affects the areas of each considered region and all the 
measured quantities we excluded from our analysis those groups whose spectroscopic and photometric redshift's 
estimates differ more than $3\epsilon(\overline{z}_{\rm phot})$\footnote{The number of discordant redshift represents 
an upper limit to the number of unconfirmed groups which is affected by the low spectroscopic completeness of SHKGs. 
While a few more groups may be recovered including galaxies with accordant spectroscopic redshifts (from SDSS or 
literature) not in the original Stoll et al. compilation, the small number ($<10$ per cent) will not significantly 
affect our results while introducing structures which do not necessarily agree with the original sample.}.

Finally we used Sloan dereddened model magnitudes k-corrected using \citet{Fukugita-1995} models, to produce for 
each group colour--magnitude diagrams for galaxies within $500\ {\rm kpc}$ and for which 
$|z_{{\rm phot}}-\overline{z}_{{\rm spec}}|\leq 3\epsilon(z_{{\rm phot}})$. We stress that even when the number 
of available spectroscopic redshifts per group is low, the confirmation of the redshift of a group obtained through 
the use of photometric redshifts (necessary to calibrate the Red Sequence), assures an accuracy greater than the CMR 
intrinsic scatter. In Fig. (\ref{fig:Fig3}, right panel) we show the colour--magnitude diagram of the group SHK 181.

\section{GROUP PROPERTIES}
\label{sec:groups-properties}

\subsection{Richness}
\label{subsec:Richness}
\begin{figure}
\includegraphics[width=0.4\textwidth,angle=0]{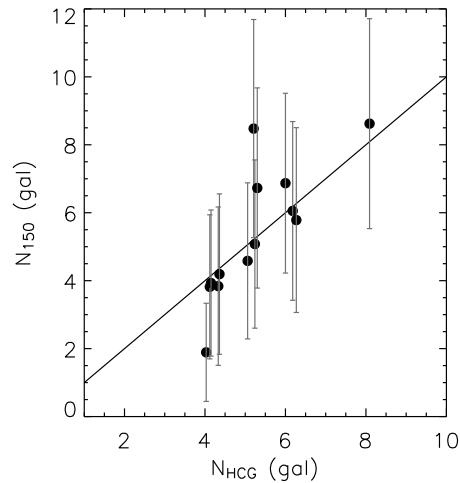} 
\caption{Comparison between the richness measurements performed for Hickson groups by 
Hickson (1982); Hickson et al. (1992) and by us. Points are slightly shifted to maximize their visibility.}
\label{fig:Fig5}
\end{figure}

We derived, for all SHK groups, richness estimates within $150$ and $500$ {\rm kpc}. 
Member galaxies have been selected as those matching simultaneously all three 
criteria defined in \S \ref{sec:themethod}, in order to allow a homogeneous comparison among the groups, sampling, 
for all of them, approximately the same magnitude range. \footnote{Note that the $r^{0}_{1}+3$ limit, adopted for 
the $N$ estimate, is always within the completeness limit of the SDSS data for our redshift range.}
The richness estimates, corrected for the local background, are reported in Table \ref{tab:Table2}.\\
In Fig. (\ref{fig:Fig5}) we present the comparison between the richness measurements performed for Hickson 
groups by \cite{Hickson-1982,Hickson-1989,Hickson-1992} and by us. We observe an overall agreement between the two 
richness estimates, consistent within the uncertainties.

\subsection{Surface Brightness Profile}
As further information we extracted the surface brightness profile of each group, selecting galaxies matching the
criteria defined in \S \ref{sec:themethod}. Starting from the group centroid, we binned the data
in annuli containing three galaxies and we calculated the average surface brightness in each one. 
The background level ($\mu_{{\rm back}}$) was evaluated by averaging all the objects between $500\ {\rm kpc}$ and 
$1\ {\rm Mpc}$. The size of a group  was arbitrarily defined as the radius where the 
surface brightness profile reaches $\mu_{{\rm back}}-2.5\log(1.5)$. Resulting values are listed in Table 
\ref{tab:Table2}. A modified estimate of the group size was also 
calculated changing the redshift constrain to $|z_{{\rm phot}}-\overline{z}_{{\rm phot}}|\leq 3\epsilon(z_{{\rm phot}})$, {\rm i.e.} using 
$\overline{z}_{{\rm phot}}$ as the reference redshift, which is more appropriate in cases where no spectroscopic 
redshift is available or our analysis shows that they are unreliable. Hereafter we'll refer to 
this last size estimate as {\it photometric}, in contrast with the first one, which will be labelled as 
{\it spectroscopic}.

\begin{figure}
\begin{center}
\includegraphics[width=0.4\textwidth,clip,angle=0]{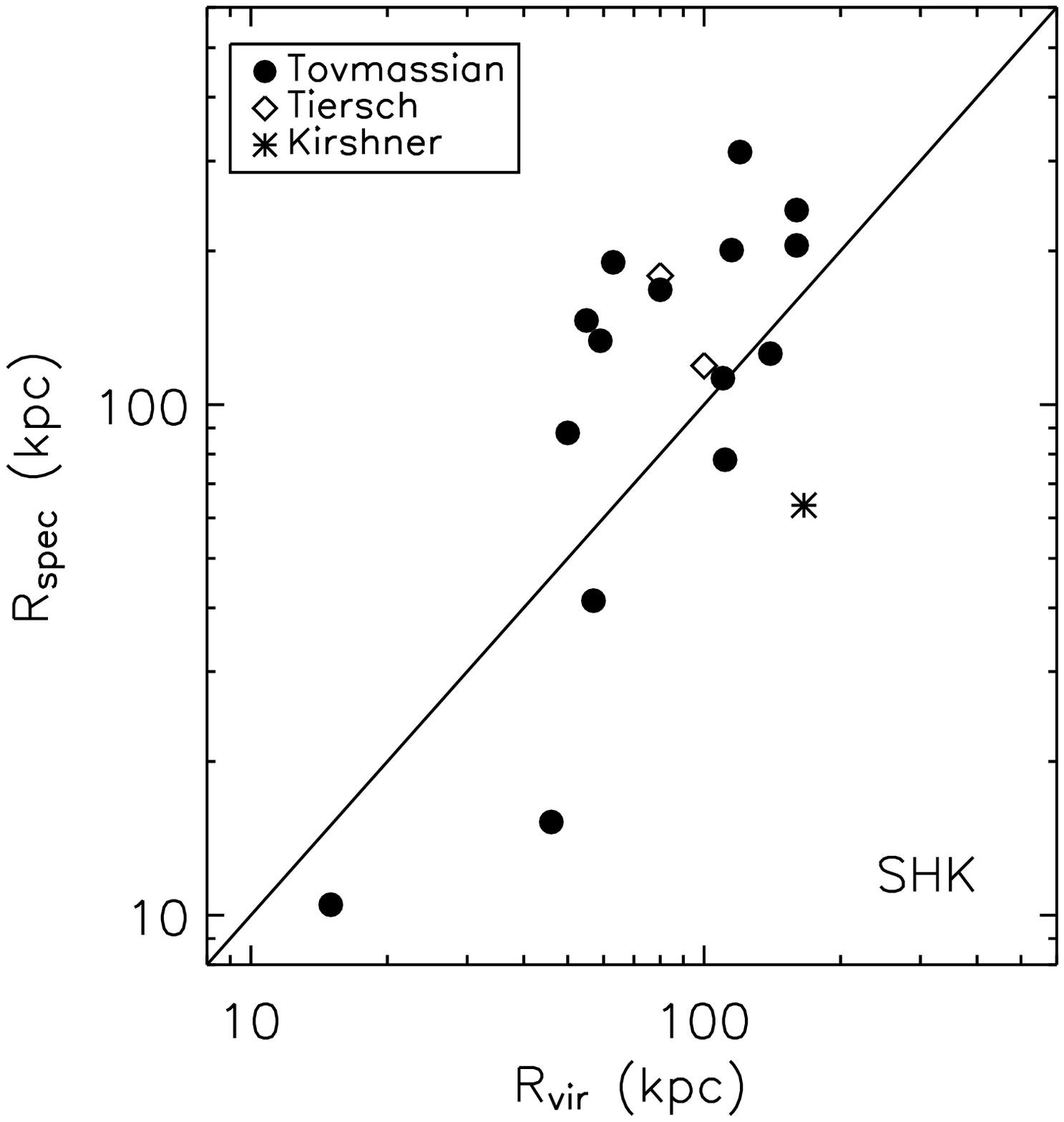} 
\hfill
\includegraphics[width=0.4\textwidth,clip,angle=0]{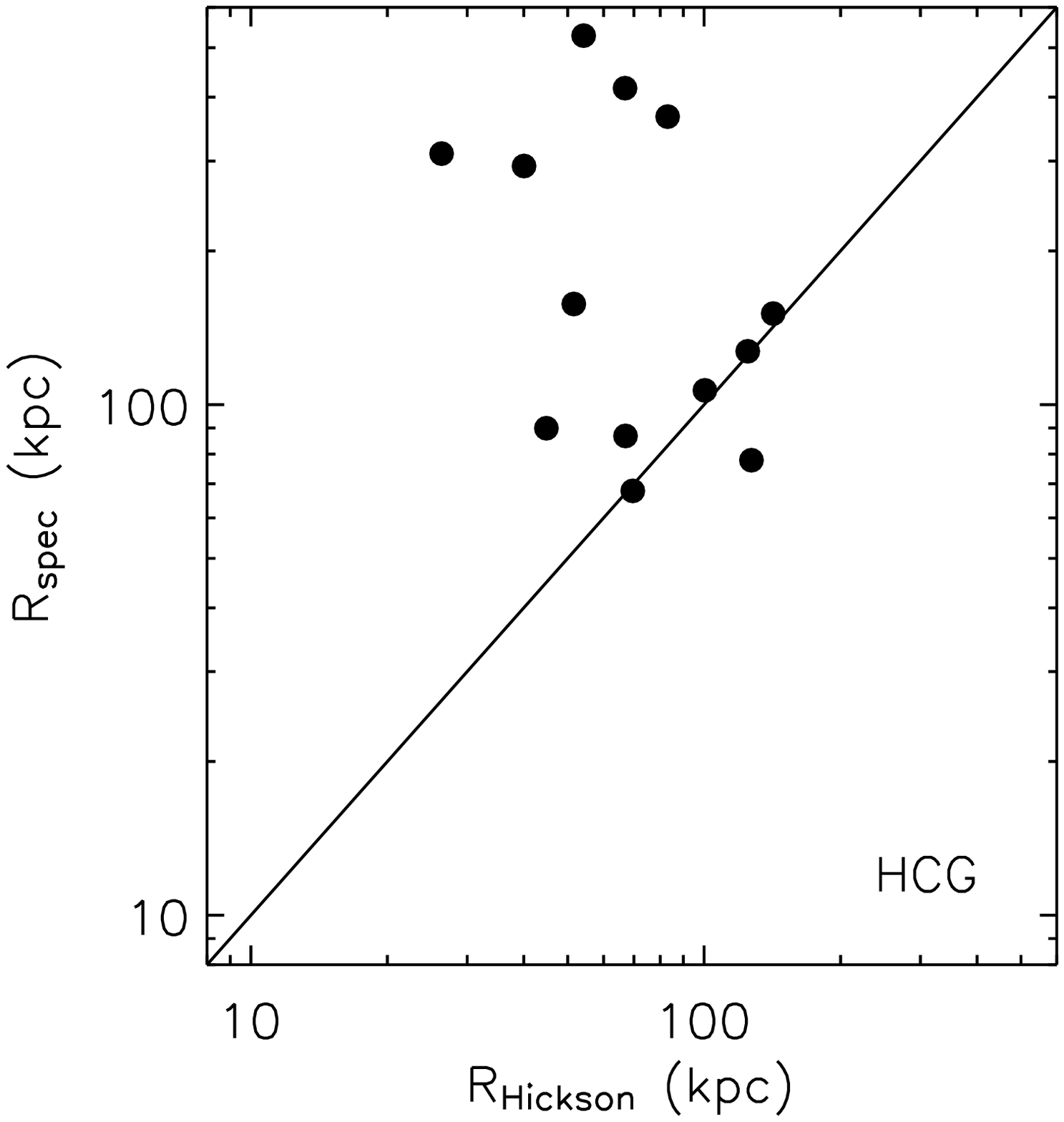} 
\caption{Comparison between our {\it spectroscopic} size estimates of the SHKs and those performed 
by Tovmassian et al. (1999-2007) (plain dots), Tiersch et al. (1994-2002) (diamonds) and \citet{Kirshner-1980}
(asterisks)(upper panel). Similarly, in the lower panel, our size estimates of the HCGs are compared with 
those performed by Hickson (1982) and Hickson et al. (1992). In both panels a logarithmic scale is used.}
\label{fig:Fig6}
\end{center}
\end{figure}

In Fig. \ref{fig:Fig6} we compare our estimates of the physical properties of both SHKs (left panel) and 
HCGs (right panel), to those available in literature. Our size estimates are consistent with the HCGs sizes for 
$\sim 50$ per cent of the sample ($\overline{R}_{{\rm spec}}=214\pm 42\ {\rm kpc}$). For the SHK sample, where sizes calculated 
according to photometric criteria are not available in literature, we compare with the virial radii of the groups 
measured by Tovmassian et al. (1999--2007) and Tiersch et al. (1994--2002). According to~\cite{Tovmassian-2005c} the 
virial radii generally don't exceed $\approx160\ {\rm kpc}$, while the mean mass-weighted radial velocity 
dispersion is $330\pm 170\ {\rm km\ s^{-1}}$ \citep{Tovmassian-2007}, ranging from $88.5$ up to $667.1\ {\rm km\ s^{-1}}$,
thus implying dynamical crossing times varying from $2.7 \times 10^{7}\ {\rm yr}$ to $1.9 \times 10^8\ {\rm yr}$ 
\citep{Tovmassian-2005c, Tovmassian-2007}, on average shorter than what found for the HCGs $\approx 2.1 \times 10^8\ {\rm yr}$ 
\citep{Hickson-1992}. While, for SHKs, we find a mild but significant correlation (correlation coefficient 
$r_{\rm corr}=0.5$ with $P_{\rm corr}=97$ per cent), with our sizes ($\overline{R}_{{\rm spec}}=142\pm 13\ {\rm kpc}$) measuring 
$\sim 50$ per cent of the virial radius, there is a large scatter due both to our purely photometric criteria as well as 
the heterogeneous accuracy of peculiar velocities available in literature, from which $r_{\rm vir}$ are calculated. 
It is worth noting that Tovmassian et al. call virial radius the deprojected harmonic mean galaxy 
separation (much smaller than the 100 over-density radius, usually defined as virial radius).

\subsection{Galaxy Morphology}
\label{subsec:Morphology}
\begin{figure}
\begin{center}
\includegraphics[width=0.4\textwidth,angle=0]{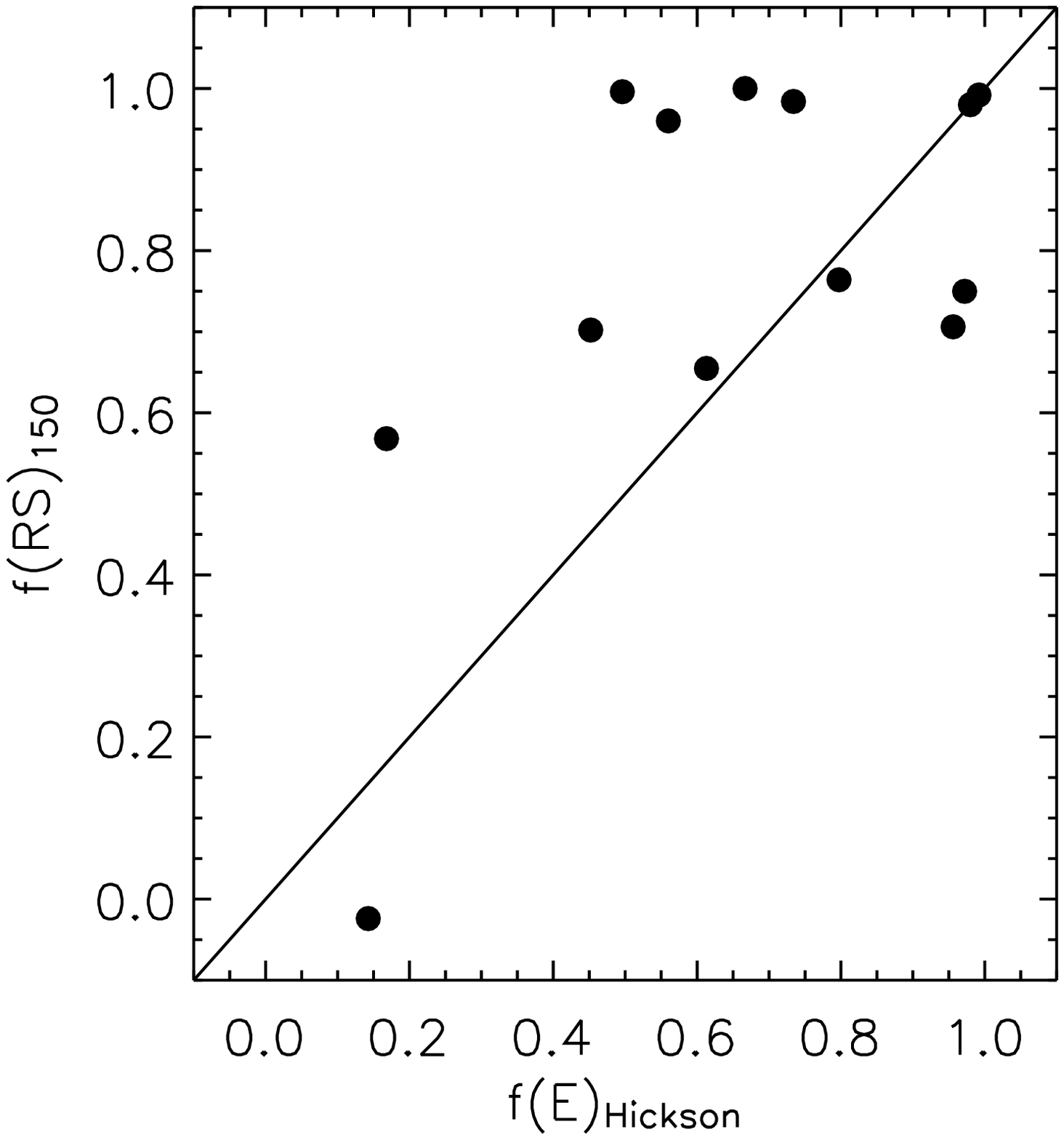}
 \caption{Comparison between the early-type fraction measurements performed for Hickson groups by 
 Hickson et al. (1989) and by us.}
\label{fig:Fig7}
\end{center}
\end{figure}

We have used as morphological indicator the distance from the {\it g-i} vs {\it r} red--sequence.
We utilised the models by T.Kodama (see \citealp{Kodama-1997} for the original paper) in the SDSS bands for
several galaxy formation redshifts, whose slopes were in accordance with our CMR points. We considered as 
early-type galaxies those within 0.15 mag from the best--fitting model in our colour--magnitude diagrams 
(hereafter Red Sequence galaxies - RS galaxies). \\
Among the photometric SDSS parameters calculated by the photometric pipeline PHOTO \citep{Lupton-2001}, concentration 
index and profile likelihood are sensitive to galaxies morphology \citep{Shimasaku-2001, Strateva-2001}. As it 
is an indicator of recent star formation, the {\it u-r} colour is an alternative parameter correlated 
with galaxies morphology, as suggested in the work by \cite{Strateva-2001}. This study indicates that galaxies 
have a bimodal {\it u-r} colour distribution corresponding to early (E, SO and Sa) and late (Sb, Sc, and Irr) 
morphological types, which can be clearly separated by a {\it u-r} colour cut of 2.22 independent on magnitude. 
For these reasons, we decided to use {\it u-r} colour, instead of concentration 
index or profile likelihood, to perform an alternative morphological classification.\\
\noindent For each group we finally derived the early-type fraction within the inner region ($f(E)_{150}$, 
$f(RS)_{150}$), the intermediate annular region ($f(E)_{{\rm env}}$, $f(RS)_{{\rm env}}$) and the local background 
($f(E)_{{\rm back}}$, $f(RS)_{{\rm back}}$), applying both methods described above to galaxies obeying to criteria 
2 and 3 defined in \S \ref{sec:themethod}. Results are listed in Table~\ref{tab:Table2}. We derived the same observables 
for HCGs and compared our early-type fractions estimates with those obtained by \cite{Hickson-1989}. This comparison 
is showed in (Fig. \ref{fig:Fig7}).  We can see that our estimates, are systematically higher that the ones reported 
by Hickson, which is not surprising since we use a photometric approach while the Hickson one is based on a purely 
visual classification.

\section{Individual properties}
\label{sec:individual}
In this section we briefly discuss the individual properties of the groups in 
our sample. \\

\noindent {\bf SHK 1}. 
A considerable and compact over-density is detected, confirmed by the presence of a well defined Red Sequence. 
The physical reality of this group is supported by a marked excess in photometric 
redshift distribution in correspondence of the spectroscopic redshift value found 
in literature. This is the richest group within our sample together with SHK 253.
 
\noindent {\bf SHK 5}. 
This group corresponds to HCG 50. The 2D galaxy over-density is confirmed by both a peak in the photometric 
redshift distribution and by the presence of an evident RS. 
This structure appears elongated in the sky plane. 

\noindent {\bf SHK 6}. 
Clear over-density in both 2D and 3D diagnostics, with a well defined RS.
The photometric redshift distribution reveals the presence of a rich background structure, at the group position,
at $z\sim 0.1$.

\noindent {\bf SHK 8}. 
According to \cite{Tovmassian-2005c} this group is contaminated by a foreground galaxy and by a star 
The central galaxies, all with consistent redshifts, appear to be interacting 
All our diagnostics fail to identify a galaxy structure; we hence reject this group from further discussion.

\noindent {\bf SHK 10}. 
Clear-cut rich group with elongated structure revealed in all diagnostics. 

\noindent {\bf SHK 11}. All the diagnostics indicate the presence of a small structure. Well defined 
Red Sequence but visible only for the few most luminous galaxies.

\noindent {\bf SHK 14}. 
Clearly defined small group with the central 6 objects all with accordant redshifts~\citep{Tovmassian-2005c}.
All our diagnostics reveal the presence of this group. Well defined Red Sequence.

\noindent {\bf SHK 19}.
Accordingly to \cite{Tovmassian-2005c}, the group is composed by 4 accordant redshift objects within a radius
of 15 {\rm kpc}. In our data we find 5 galaxies in a radius of $\sim 10\ {\rm kpc}$ and a background structure at
$z\sim 0.15$.

\noindent {\bf SHK 22}. 
Poor and dispersed structure visible in 2D diagnostics and in photometric redshift distribution. No signs of Red
Sequence.

\begin{table*}
\begin{tabular}{lcccccccccc}
\hline
\hline
 {\bf SHK} & {${\mathbf{f(RS)_{150}}}$} & {${\mathbf{f(RS)_{{\rm {\bf env}}}}}$} &  {${\mathbf{f(RS)_{{\rm {\bf back}}}}}$}& {${\mathbf{N_{500}}}$}& 
 {${\mathbf{N_{150}}}$} & {$\mathbf{N_{{\rm {\bf env}}}}$} & {${\mathbf{R_{{\rm {\bf spec}}}}}$} & {$\mathbf{N_{R_{{\rm {\bf spec}}}}}$} & 
 {${\mathbf{R_{{\rm {\bf phot}}}}}$} & {$\mathbf{N_{R_{{\rm {\bf phot}}}}}$}\\
  1           &     2	    &	  3	  &    4    &	    5	   &	    6	  &	 7	   &    8       &     9       &    10	   &	  11	 \\									      
\hline
  1	      &   0.86      &	0.44	  &   0.42  &	 24$\pm$6   &	  13$\pm$4  &	 35$\pm$8    &    63	  &    12$\pm$3   &   63      &   12$\pm$3   \\   
  5	      &   1.00      &	0.40	  &   0.38  &	 6 $\pm$3   &	  5 $\pm$2  &	 8 $\pm$6    &    145	  &    4 $\pm$2   &   145     &   4 $\pm$2   \\   
  6	      &   1.00      &	0.64	  &   0.40  &	 23$\pm$5   &	  7 $\pm$3  &	 33$\pm$7    &    30	  &    3 $\pm$2   &   30      &   3 $\pm$2   \\   
  10	      &   0.71      &	0.65	  &   0.46  &	 25$\pm$5   &	  7 $\pm$3  &	 31$\pm$7    &    302	  &    17$\pm$4   &   302     &   17$\pm$4   \\   
  11	      &   1.00      &	0.29	  &   0.10  &	 9 $\pm$4   &	  4 $\pm$2  &	 6 $\pm$5    &    269	  &    7 $\pm$3   &   224     &   8 $\pm$3   \\   
  14	      &   0.50      &	0.47	  &   0.47  &	 10$\pm$4   &	  8 $\pm$3  &	 4 $\pm$5    &    41	  &    5 $\pm$2   &   41      &   5 $\pm$2   \\   
  19	      &   0.80      &	0.47	  &   0.46  &	 17$\pm$5   &	  5 $\pm$2  &	 46$\pm$6    &    10	  &    2 $\pm$1   &   ---     &     -----    \\
  22	      &   0.20      &	0.46	  &   0.39  &	 4 $\pm$3   &	  5 $\pm$2  &	 8 $\pm$5    &    201	  &    5 $\pm$2   &   201     &   5 $\pm$2   \\   
  31	      &   1.00      &	0.93	  &   0.86  &	 11$\pm$4   &	  6 $\pm$2  &	 5 $\pm$4    &    312	  &    10$\pm$3   &   312     &   10$\pm$3   \\   
  54	      &   1.00      &	0.57	  &   0.20  &	 17$\pm$5   &	  5 $\pm$2  &	 21$\pm$7    &    331	  &    17$\pm$4   &   434     &   15$\pm$4   \\   
  55	      &   0.86      &	0.19	  &   0.08  &	 7 $\pm$3   &	  7 $\pm$3  &	 3 $\pm$5    &    88	  &    5 $\pm$2   &   88      &   5 $\pm$2   \\   
  57	      &   0.60      &	0.54	  &   0.29  &	 17$\pm$5   &	  10$\pm$3  &	 21$\pm$6    &    233	  &    10$\pm$3   &   233     &   10$\pm$3   \\   
  63	      &   0.67      &	0.25	  &   0.03  &	 1 $\pm$2   &	  3 $\pm$2  &	 -3$\pm$2    &    52	  &    3 $\pm$2   &   215     &   4 $\pm$2   \\   
  65	      &   0.00      &	0.33	  &   0.10  &	 36$\pm$7   &	  7 $\pm$3  &	 37$\pm$10   &    110	  &    5 $\pm$2   &   93      &   2 $\pm$2   \\   
  74	      &   0.17      &	0.49	  &   0.49  &	 20$\pm$6   &	  5 $\pm$2  &	 47$\pm$10   &    240	  &    18$\pm$5   &   252     &   23$\pm$5   \\   
  95	      &   1.00      &	0.64	  &   0.07  &	 3 $\pm$7   &	  4 $\pm$2  &	 -2$\pm$4    &    32	  &    2 $\pm$1   &   32      &   2 $\pm$1   \\   
  120	      &   0.80      &	0.37	  &   0.38  &	 10$\pm$4   &	  4 $\pm$2  &	 6 $\pm$6    &    15	  &    3 $\pm$2   &   96      &   8 $\pm$3   \\   
  123	      &   0.57      &	0.37	  &   0.08  &	 12$\pm$4   &	  6 $\pm$3  &	 0 $\pm$5    &    229	  &    7 $\pm$3   &   229     &   7 $\pm$3   \\   
  128	      &   0.80      &	0.29	  &   0.06  &	 4 $\pm$3   &	  5 $\pm$2  &	 1 $\pm$4    &    38	  &    3 $\pm$2   &   38      &   3 $\pm$2   \\   
  154	      &   0.89      &	0.75	  &   0.75  &	 30$\pm$6   &	  8 $\pm$3  &	 46$\pm$9    &    119	  &    9 $\pm$3   &   232     &   8 $\pm$3   \\   
  181	      &   0.90      &	0.53	  &   0.49  &	 24$\pm$5   &	  10$\pm$3  &	 27$\pm$6    &    168	  &    12$\pm$3   &   150     &   11$\pm$3   \\   
  186	      &   0.67      &	0.31	  &   0.11  &	 13$\pm$4   &	  6 $\pm$2  &	 3 $\pm$4    &    59	  &    6 $\pm$2   &   59      &   6 $\pm$2   \\   
  188	      &   1.00      &	0.64	  &   0.64  &	 7 $\pm$3   &	  6 $\pm$2  &	 -1$\pm$4    &    126	  &    6 $\pm$2   &   126     &   6 $\pm$2   \\   
  191	      &   1.00      &	0.67	  &   0.64  &	 18$\pm$5   &	  11$\pm$3  &	 24$\pm$6    &    190	  &    15$\pm$4   &   266     &   17$\pm$4   \\   
  205	      &   1.00      &	0.39	  &   0.40  &	 16$\pm$4   &	  4 $\pm$2  &	 16$\pm$6    &    178	  &    4 $\pm$2   &   178     &   4 $\pm$2   \\   
  213	      &   0.80      &	0.60	  &   0.60  &	 7 $\pm$3   &	  6 $\pm$2  &	 -2$\pm$4    &    133	  &    5 $\pm$2   &   212     &   4 $\pm$2   \\   
  218	      &   0.50      &	0.25	  &   0.25  &	 2 $\pm$3   &	  6 $\pm$3  &	 -9$\pm$5    &    259	  &    6 $\pm$3   &   136     &   8 $\pm$3   \\   
  223	      &   1.00      &	0.55	  &   0.53  &	 21$\pm$5   &	  9 $\pm$3  &	 28$\pm$7    &    146	  &    8 $\pm$3   &   210     &   11$\pm$3   \\   
  231	      &   0.80      &	0.30	  &   0.06  &	 13$\pm$4   &	  4 $\pm$2  &	 4 $\pm$5    &    178	  &    4 $\pm$2   &   178     &   4 $\pm$2   \\   
  237	      &   1.00      &	0.42	  &   0.40  &	 9 $\pm$4   &	  3 $\pm$2  &	 21$\pm$7    &    194	  &    4 $\pm$2   &   35      &   2 $\pm$1   \\   
  245	      &   0.57      &	0.62	  &   0.64  &	 18$\pm$5   &	  9 $\pm$3  &	 35$\pm$7    &    133	  &    8 $\pm$3   &   133     &   8 $\pm$3   \\   
  251	      &   0.57      &	0.42	  &   0.42  &	 6 $\pm$4   &	  6 $\pm$3  &	 -6$\pm$5    &    88	  &    5 $\pm$2   &   88      &   5 $\pm$2   \\   
  253	      &   0.92      &	0.38	  &   0.50  &	 15$\pm$4   &	  13$\pm$4  &	 13$\pm$5    &    90	  &    12$\pm$3   &   90      &   12$\pm$3   \\   
  254	      &   0.80      &	0.21	  &   0.04  &	 6 $\pm$4   &	  4 $\pm$2  &	 -2$\pm$6    &    57	  &    3 $\pm$2   &   57      &   3 $\pm$2   \\   
  344	      &   0.78      &	0.29	  &   0.39  &	 10$\pm$4   &	  8 $\pm$3  &	 -4$\pm$5    &    113	  &    8 $\pm$3   &   113     &   8 $\pm$3   \\   
  346	      &   0.78      &	0.72	  &   0.64  &	 16$\pm$5   &	  8 $\pm$3  &	 -4$\pm$5    &    134	  &    7 $\pm$3   &   151     &   7 $\pm$3   \\   
  348	      &   1.00      &	0.67	  &   0.67  &	 16$\pm$4   &	  6 $\pm$2  &	 12$\pm$5    &    205	  &    8 $\pm$3   &   146     &   6 $\pm$2   \\   
  351	      &   0.62      &	0.40	  &   0.16  &	 16$\pm$5   &	  9 $\pm$3  &	 19$\pm$6    &    298	  &    10$\pm$3   &   298     &   10$\pm$3   \\   
  352	      &   1.00      &	0.80	  &   0.75  &	 28$\pm$6   &	  9 $\pm$3  &	 33$\pm$7    &    92	  &    8 $\pm$3   &   92      &   8 $\pm$3   \\   
  355	      &   1.00      &	0.44	  &   0.41  &	 4 $\pm$3   &	  3 $\pm$2  &	 -4$\pm$5    &    29	  &    3 $\pm$2   &   29      &   3 $\pm$2   \\   
  357	      &   1.00      &	0.77	  &   0.66  &	 34$\pm$6   &	  10$\pm$3  &	 43$\pm$7    &    211	  &    13$\pm$4   &   211     &   13$\pm$4   \\   
  358	      &   1.00      &	0.29	  &   0.07  &	 5 $\pm$3   &	  6 $\pm$2  &	 1 $\pm$3    &    86	  &    5 $\pm$2   &   86      &   5 $\pm$2   \\   
  360	      &   1.00      &	0.81	  &   0.87  &	 23$\pm$5   &	  11$\pm$3  &	 34$\pm$6    &    179	  &    15$\pm$4   &   179     &   15$\pm$4   \\   
  371	      &   0.50      &	0.32	  &   0.25  &	 2 $\pm$3   &	  5 $\pm$2  &	 -1$\pm$5    &    93	  &    5 $\pm$2   &   93      &   5 $\pm$2   \\   
  376	      &   0.86      &	0.79	  &   0.78  &	 10$\pm$4   &	  7 $\pm$3  &	 16$\pm$6    &    78	  &    6 $\pm$2   &   93      &   5 $\pm$2   \\   
\hline														
\end{tabular}								  					
\caption{{\bf Groups Properties.}													
Column 1: SHK identification number;											
column 2: Red Sequence galaxies fraction within the inner region;  		    									
column 3: Red Sequence galaxies fraction within the environment region; 	    										
column 4: Red Sequence galaxies fraction within local background;
column 5: richness derived within 500 {\rm kpc} according to the criteria defined in
\S \ref{sec:themethod};
column 6: richness derived within 150 {\rm kpc} according to the criteria defined in 
\S \ref{sec:themethod}; 
column 7: richness derived within the environment region according to the criteria 2 and 3 defined in 
\S \ref{sec:themethod}; 
column 8: group size spectroscopic estimate given in {\rm kpc};
column 9: richness derived within the group size {\it spectroscopic} estimate according to the 
criteria 2 and 3 defined in \S \ref{sec:themethod};
column 10: group size photometric estimate given in {\rm kpc};
column 11: richness derived within the group size {\it photometric} estimate according to the 
criterion 2 defined in \S \ref{sec:themethod} and the photometric redshift selection: 
	  $|z_{{\rm phot}}-\overline{z}_{{\rm phot}}|\leq 3\epsilon(z_{{\rm phot}})$.}
\label{tab:Table2}	  
\end{table*}

\noindent {\bf SHK 29}. 
The spectroscopic redshift in literature refers to only one object, probably of foreground. No signs of 
over-density or of a RS are detected in correspondence of the literature spectroscopic redshift estimate. 
In the photometric redshift distribution a well defined structure is evident at $z_{{\rm phot}}\sim 0.17$. The group 
seems to belong to an high redshift structure. 
Since  $\Delta z>3\mathbf{\epsilon(\overline{z}_{{\rm phot}})}$ we reject this group from further discussion.

\noindent {\bf SHK 31}. 
Loose and well defined group, visible in all diagnostics. RS is clearly visible.

\noindent {\bf SHK 54}. 
Complex structure: two clumps, of comparable richness at the same redshift. 
There is a well defined RS. 

\noindent {\bf SHK 55}. Well defined compact and isolated group. 

\noindent {\bf SHK 57}. Well defined high redshift group evident in all diagnostics.

\noindent {\bf SHK 60}. Poor compact group with larger and looser structure in the background. The spectroscopic and 
photometric redshift estimates differ by more than $3\epsilon(\overline{z}_{{\rm phot}})$. We reject this group 
from further discussion.  

\noindent {\bf SHK 63}. 
This is an isolated triplet of galaxies at redshift consistent with the spectroscopic one found in literature,
whose 2D over-density is raised by the superposition of other two foreground galaxies. 

\noindent {\bf SHK 65}. No evidence for structure in spectroscopic redshifts distribution. In the other diagnostics we 
observe a large and loose structure with a quite extended over-density in photometric redshifts and a very poor RS. 

\noindent {\bf SHK 70}. Very poor structure partially superposed with with a structure at $z\sim 0.15$. The low
S/N ratio makes it hard to detect this group with our diagnostic tools. We exclude it from the final sample.

\noindent {\bf SHK 74}. 
\cite{Tovmassian-2005a} report the existence of two distinct groups at two different redshifts. We don't find 
hints of the presence of two distinct structures in our photometric redshift distribution, but of only one, loose
structure at $z=0.12$.

\noindent {\bf SHK 95}. Small excess of spectroscopic redshifts. Only moderate evidence of
a structure in the other diagnostics. We interpreted it as a poor group at low redshift.

\noindent {\bf SHK 96}. 
No diagnostic reveals the presence of a group. Its detection was hence most probably due to a chance galaxy 
superposition. We reject it from the sample.

\noindent {\bf SHK 104}. Poor group, possibly isolated presenting no excess in the spectroscopic redshift distribution.
Evidence of a structure in the 2D diagnostics, but only moderate in the 3D ones. No presence of a Red Sequence. 
Mismatch between spectroscopic and photometric redshifts greater than $3\epsilon(\overline{z}_{{\rm phot}})$. We 
reject it from the sample.

\noindent {\bf SHK 120}. Both density maps and photometric redshift distributions reveal the presence of a rich, 
elongated structure projected against a looser and more distant one. Slightly defined RS.  

\noindent {\bf SHK 123}. 
This low multiplicity group shows a pronounced peak in the photometric redshift distribution and a marginally 
defined RS. Two different groups of objects are present, maybe at similar redshifts and possibly belonging to 
the same structure. 

\noindent {\bf SHK 128}. Compact and isolated group of low multiplicity. 

\noindent {\bf SHK 152}. No diagnostics present hints of the presence of a structure. 
$\Delta z>3\mathbf{\epsilon(\overline{z}_{{\rm phot}})}$.Rejected from final sample.

\noindent {\bf SHK 154}. 
The group has been studied in detail by \cite{Tiersch-2002} who found 5 galaxies with accordant redshifts. 
They also detected signs of interactions among the members as well as an extended halo surrounding the group. In 
addition, they find that some galaxies which appear to be projected on the main group have discordant redshifts. 
We find a strong 2D over-density with well defined RS and a strong excess of photometric redshifts. The presence 
of a secondary peak at $\sim 270\ {\rm kpc}$ in the numeric surface density radial profile could indicate the 
presence of a second structure at higher redshift which is also confirmed by a double peak in the photometric 
redshift distribution.

\noindent {\bf SHK 181}. 
Rich group, with very well defined over-density and elongated appearance.
Its physical nature was proved by detailed photometric and spectroscopic studies by 
\cite{Fasano-1994} and \cite{Tovmassian-2004}. \cite{Tovmassian-2004} measure a rather low spatial density compared 
the other SHKs. Our data show the presence of a well defined RS and of a pronounced excess in photometric 
redshift distribution (Figs. \ref{fig:Fig1},\ref{fig:Fig2} and \ref{fig:Fig3}).

\noindent {\bf SHK 184}. The mean spectroscopic and photometric redshifts of the group differ significantly and 
there is no clear evidence of the presence of a structure consistent with the spectroscopic value. Rejected

\noindent {\bf SHK 186}. Compact and elongated (chain-like) structure. Visible in all diagnostics.

\noindent {\bf SHK 188}. 
Rich group slightly off-centred with respect to position listed in Stoll's catalogue. 
Well defined RS.

\noindent {\bf SHK 191}. 
This group is the core of Abell cluster A1097~\citep{Tovmassian-2005b}. The excess in the photometric redshift 
distribution is highly evident and the RS is clearly defined.

\noindent {\bf SHK 202}. 
Even though optical and X-Ray band studies \citep{Kodaira-1988,Kodaira-1990, Takahashi-2000,Takahashi-2001} confirm the physical
reality of this group, because of its closeness, resulting in a too low S/N ratio, our diagnostic tools fail to
detect it as a statistic excess. Rejected from further analysis.

\noindent {\bf SHK 205}. 
Loose group of low richness settling on the expected RS.
The average photometric and spectroscopic redshifts slightly disagree. 

\noindent {\bf SHK 213}. 
Loose group showing a clumpy core, a moderate over-density and a well defined RS. 
The photometric redshift distribution presents a well defined excess at a redshift slightly higher than the 
assumed one. 

\noindent {\bf SHK 218}. 
Well defined group confirmed by RS and excess of photometric redshifts. 
There is evidence for a second structure in the background  at redshift $z\sim 0.18$. 

\noindent {\bf SHK 223}. 
Loose and rich group, having a well defined RS. It shows slightly off-centred substructures. 

\noindent {\bf SHK 229}. $\Delta z>3\mathbf{\epsilon(\overline{z}_{{\rm phot}})}$. Our diagnostics don't find any hints
of the presence of a structure. Maybe this is due to the fact that the spectroscopic redshift estimate found in
literature referred to only one object. Rejected from further analysis.

\noindent {\bf SHK 231}. Compact, elongated structure of low multiplicity. 
Visible in all diagnostics.

\noindent {\bf SHK 237}.
Nearby poor group. Moderate hints in our diagnostics of the presence of a structure.

\noindent {\bf SHK 245}. 
Well defined over-density of elongated appearance.
The RS is well defined and both the spectroscopic and photometric redshift 
distributions hint to the existence of a second background structure. 

\noindent {\bf SHK 248}. $\Delta z>3\mathbf{\epsilon(\overline{z}_{{\rm phot}})}$, hence rejected from further analysis.
Our 2D diagnostics found hints of the over-density, but photometric redshift distribution shows that this group is
closer than indicated by the spectroscopic redshift found in literature. 

\noindent {\bf SHK 251}. 
Well defined but not very strong over-density which extends well beyond the central condensation. 
RS detected but not conclusive. The background seems to indicate the presence of a rich structure at the edges 
of the field. 

\noindent {\bf SHK 253}. 
Compact, isolated structure elongated in shape. Strong excess of photometric redshifts. Well defined red sequence. 

\noindent {\bf SHK 254}. 
Moderate hints of the presence of a structure (triplet) in the inner $150\ {\rm kpc}$. This structure seems to
extend further out this radial distance, including other 4 galaxies.

\noindent {\bf SHK 258}. No diagnostics present hints of the presence of a structure. Optical group. Rejected.

\noindent {\bf SHK 344}. 
\cite{Tovmassian-2004} performed a detailed analysis of this group, also revealing moderate signs of 
interaction among some of the galaxies in the core. In our data, we find a well defined over-density with a 
clearly visible RS and a second structure in the background. 

\noindent {\bf SHK 346}. 
Strong over-density clearly visible in all diagnostics. Evidence for a background compact substructure. 

\noindent {\bf SHK 348}. 
Loose group, clearly visible in all diagnostics.

\noindent {\bf SHK 351}. Nearby, extended group. A pronounced excess in photometric redshifts and a well defined CMR 
are evident. A background extended structure is clearly visible.

\noindent {\bf SHK 352}. 
Rich and well defined group of compact appearance. The RS is clearly visible and there is a well pronounced 
excess of both spectroscopic and photometric redshifts.

\noindent {\bf SHK 355}. 
Small compact and very well defined group of galaxies. 
Its CMR is evident.

\noindent {\bf SHK 357}.  
It is a rich and compact group clearly detected as both an excess in photometric redshift distribution and a 
well defined RS.

\noindent {\bf SHK 358}. Nearby group. Strong excess in the photometric redshift distribution in correspondence of the 
mean spectroscopic redshift measure. Well defined RS. 

\noindent {\bf SHK 359}. 
Poor, nearby structure clearly overlapping with a background one. 
$\Delta z>3\mathbf{\epsilon(\overline{z}_{{\rm phot}})}$. Rejected.

\noindent {\bf SHK 360}. 
This group is the central part of the Abell cluster 2113 and it has been studied in detail 
by \cite{Tiersch-2002} who proved that the central galaxies have accordant redshifts 
and that the brightest objects are interacting. The group is almost entirely composed of early-type (E/S0) 
galaxies. In our data, the group is one of the the richest groups in the sample with a compact core and a very 
well defined RS. A significant excess is found out to $1\ {\rm Mpc}$. The photometric redshift excess appears to 
be slightly off-centred with respect to what is listed in both Stoll's catalogue and in \cite{Tiersch-2002}. 

\noindent {\bf SHK 371}. 
Poor, compact and well defined group, elongated in shape. 

\noindent {\bf SHK 376}. 
According to \cite{Tovmassian-2003b} this is a peculiar SHK group, exclusively composed by spirals.
We refer to this paper for a detailed discussion of membership and interactions.
Our data confirm that the group is embedded within a larger structure extending out 
to $\sim 150\ {\rm kpc}$, but not the morphology of the galaxies, which, in partial contrast with 
\cite{Tovmassian-2003b}, we found to be almost all early-type.\\

We excluded from further analysis the groups having $\Delta z>3\mathbf{\epsilon(\overline{z}_{{\rm phot}})}$ and 
for which the diagnostics we used didn't show hints of the presence of a physical structure. 
According to the above discussion we excluded the groups: SHK 8, SHK 29, SHK 60, SHK 70, SHK 96, SHK 104, SHK 152,
SHK 184, SHK 202, SHK 229, SHK 248, SHK 258 and SHK 359, which leaves us with a homogeneous sample of 45 SHKs.

\begin{figure*}
\includegraphics[width=0.48\textwidth,clip,angle=0]{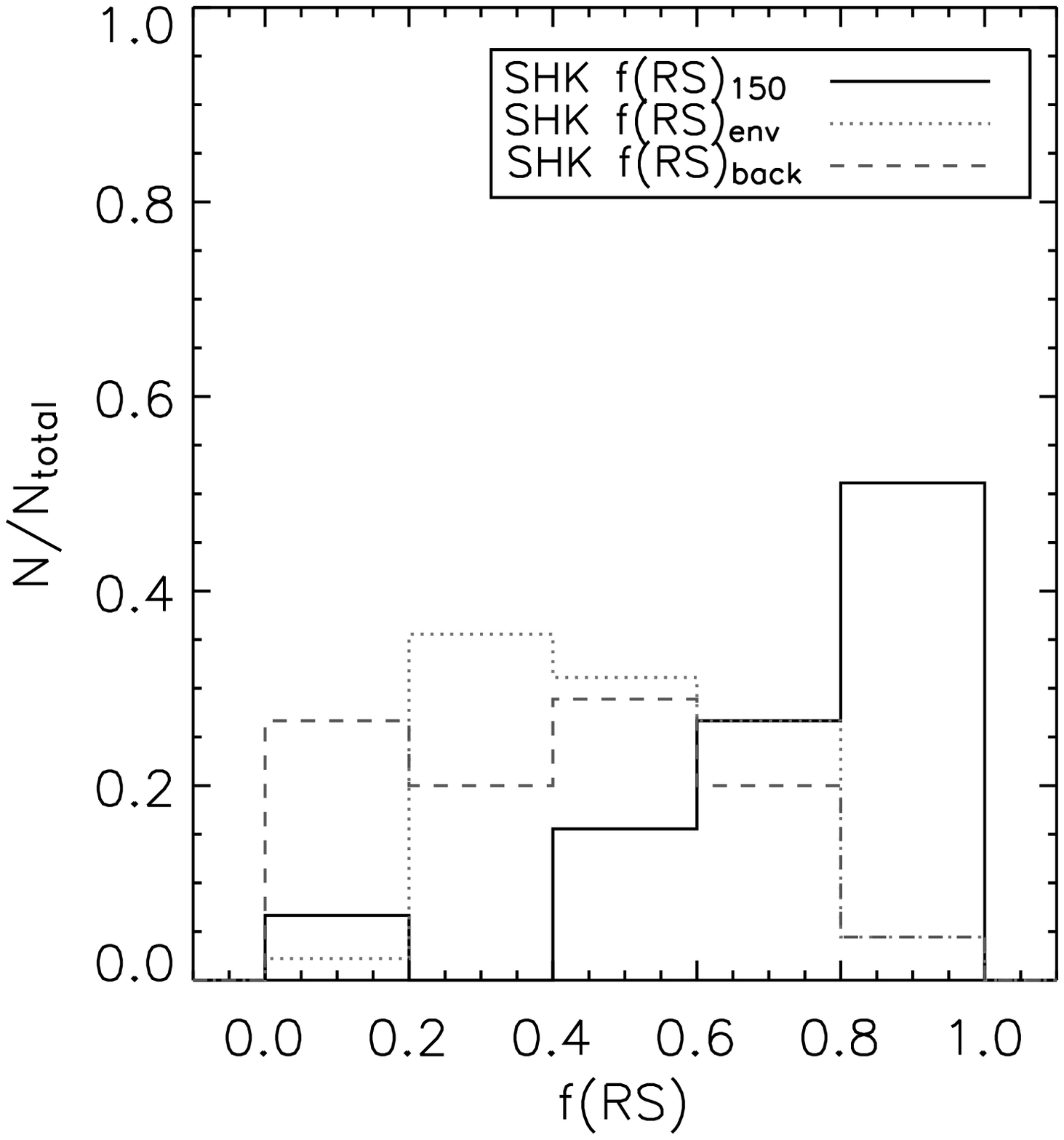} 
\includegraphics[width=0.48\textwidth,clip,angle=0]{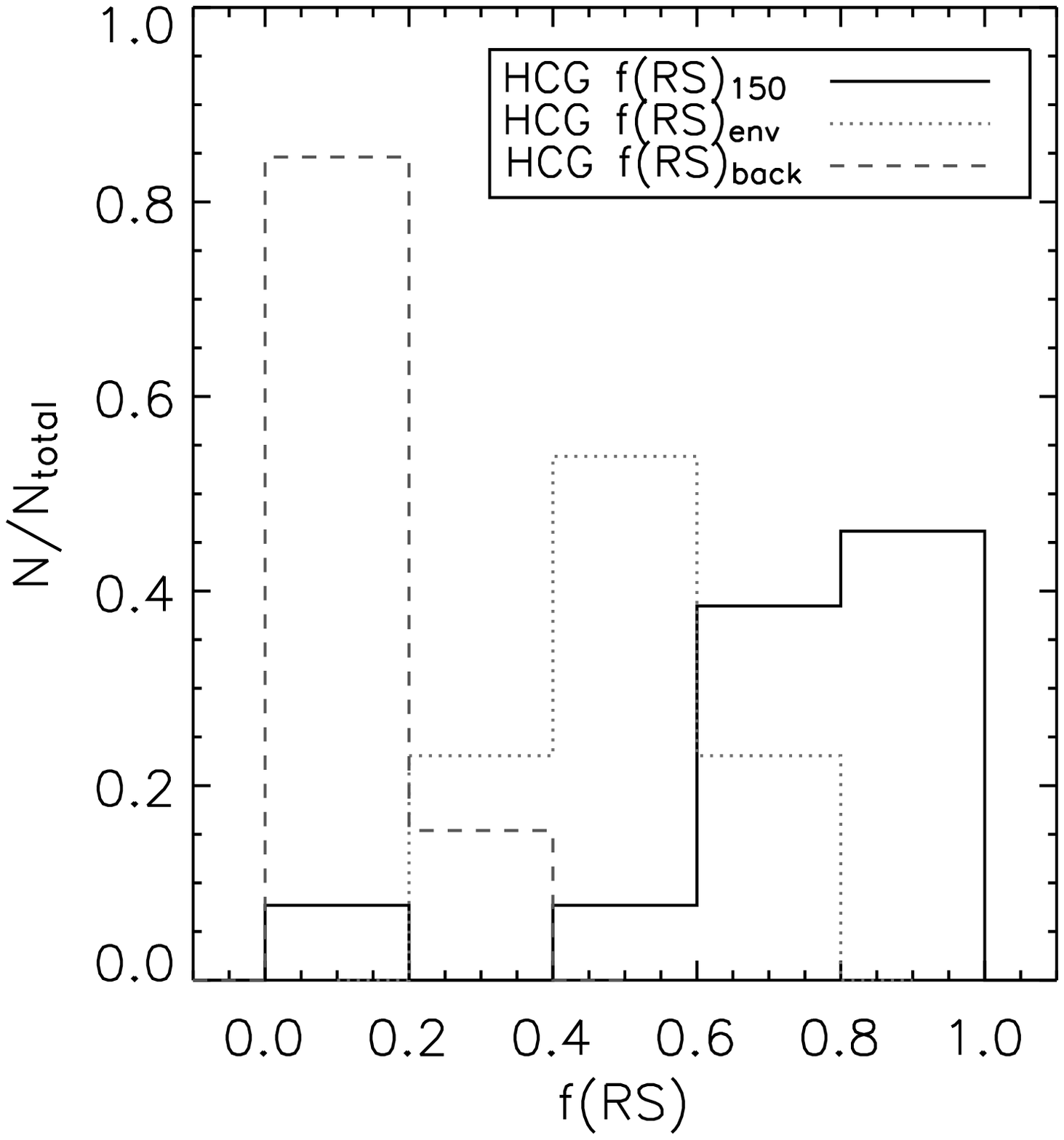} 
\caption{Left panel: normalised red--sequence galaxies fraction distributions for SHKs' inner regions (solid line), 
environment (dotted line) and local background (dashed line).
Right panel: normalised red--sequence galaxies fraction distributions for HCGs' inner regions (solid line), 
environment (dotted line) and local background (dashed line).}
\label{fig:Fig8}
\end{figure*}

\section{Results}
\label{sec:Results}
The analysis performed in the previous sections shows that about $77$ per cent of SHK groups
are likely real structures, with richness ranging from 3 to 13 galaxies, characterized by the presence 
of a projected spatial over-density, an excess in the photometric redshift distribution and 
compatible mean spectroscopic and photometric redshifts. Furthermore, most of them present a well-defined CMR 
consistent with the predictions of \cite{Bernardi-2003}. 
 For the remaining $23$ per cent we can identify several cases: 
i) the group is not real, but likely due to projection effects (SHK 96, 258); 
ii) the group is either a very poor structure or located at very low redshift ($<0.03$) and largely contaminated by background galaxies so that, due to the low S/N ratio, our approach does not allow a reliable detection (SHK 8,
104, 202); 
iii) the mean spectroscopic and photometric redshifts of the group differ significantly and/or there is no clear 
evidence of the presence of a structure consistent with the spectroscopic value (SHK 29, 70, 152, 184, 229, 248, 359);
iv) there is more than one excess in the photometric redshift distribution, revealing the 
superposition of different structures at different distances (SHK 60).\\
For the last two cases the group properties could be correctly estimated, adopting 
the photometric redshift estimate. This is though beyond of the purpose of this paper, in which we concentrate on 
estimating all properties in a homogeneous way.

We applied the above analysis also to the sub-sample of HCGs discussed in Sec.\ref{sec:themethod}. We find that 
$87$ per cent of this sub-sample is classified as a real group of galaxies according to our criteria. 
For comparison with published group catalogues, we cross-correlated the whole sample of 214 Shakhbazyan's groups 
contained within the SDSS DR5, to catalogues of groups of galaxies such as those of \cite{Lee-2004, Berlind-2006, Tago-2008}
and  \cite[UZC]{Focardi-2002}. 
Surprisingly, using a matching radius of $r=3'$ ($> 550$ {\rm kpc} for our redshift range), we find a small overlap with the above 
catalogues. The catalogues by Tago et al. and Berlind et al. contain only $53$ and $10$ groups respectively that 
are consistent with Shakhbazyan groups, while no SHK group is found in the remaining catalogues. This result shows 
that the current generation of group finding algorithms are largely missing this class of poor structures, mainly 
because they are tailored to find isolated structures, but also because the spectroscopic samples they use are 
highly incomplete for such groups. This result has strong consequences for cosmological studies where the 
availability of catalogues of cosmic structures of low multiplicity is essential.

To better understand the environment inhabited by SHKs, we looked for the presence of nearby clusters cross
correlating their positions with those of Abell's and Zwicky's galaxies clusters. Within a radius of $6'$ we found 
12 Abell and 9 Zwicky clusters. Within the 45 SHKs final sample, only 9 groups (SHK 6, SHK 22, SHK 54, SHK 65, 
SHK 120, SHK 154, SHK 181, SHK 191, SHK 357, SHK 360)  could dwell inside or reside near Abell or Zwicky clusters.

\subsection{Global Properties}
\label{subsec:global}
Using the individual properties listed in Sect. \ref{sec:individual} together with
the parameters in Table (\ref{tab:Table2}), we can now study the global optical properties of the groups in our 
sample.\\
From the morphological point of view, several studies showed that SHKs, as expected from their 
selection criteria, are on average rich in early-type (E/S0) galaxies ($77$ per cent against $51$ per cent in HCGs and $40$ 
per cent in the field), which are on average very red (${\it B-V}\geq 1.0$ and ${\it R-K} = 2.9 \pm 0.6$ 
\citep{Tovmassian-2007}).

\begin{figure}
\begin{center}
\includegraphics[width=0.45\textwidth,clip,angle=0]{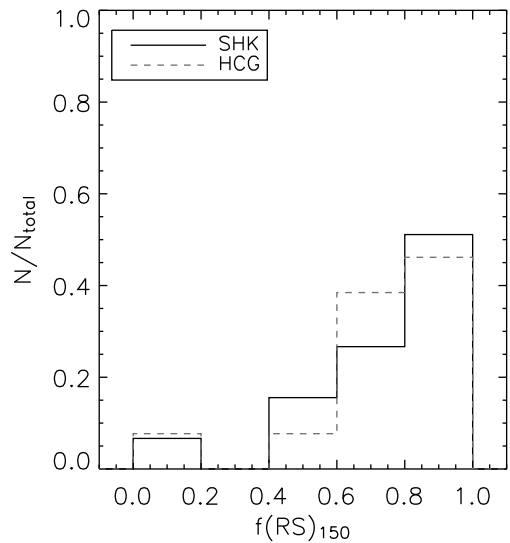}
\caption{Normalised inner ($R<150\ {\rm kpc}$) red--sequence galaxies fraction distributions of the SHKs 
(solid line), and of the HCGs (dotted line)}
\label{fig:Fig9}
\end{center}
\end{figure}

According to our photometric and CMR morphological classification, $>90$ per cent and $>75$ per cent of SHKs have 
$f(E)_{150}>0.6$ and $f(RS)_{150}>0.6$ respectively.
The overabundance of early-type galaxies is however confined to the group core since when we consider 
the early-type fraction in the inner, intermediate and local background regions (Fig. \ref{fig:Fig8}), we note 
that the dominant morphological type changes when moving outwards. This trend characterizes both SHKs 
(Fig. \ref{fig:Fig8}, left panel) and HCGs (Fig. \ref{fig:Fig8}, right panel), and reflects, to some extent, 
the morphology--density relation \citep{Dressler-1980}.
When compared with our sub-sample of HCGs, we find that the $f(E)_{150}$ and $f(RS)_{150}$ distributions are skewed 
toward systematically higher early-type fractions ({\rm e.g.}, see Fig. \ref{fig:Fig9}). According to our classification, 
$\sim 70$ per cent of the SHKs we analysed have $f(E)_{150}\geq 0.8$ against $\sim 55$ per cent for the HCG sample, 
even though, given the small HCG sample size (13 HCGs with $z_{{\rm spec}} \geq 0.03$), this difference is very not 
significant, (K-S probability of $80$ per cent). As discussed in \S \ref{subsec:Morphology}, our classification 
methods differ from the purely morphological one adopted by~\cite{Hickson-1982} and \cite{Hickson-1992}. According to 
Hickson's criteria, the sub-sample included in the present work shows an $f(E)\sim 40$ per cent, while for the 
complete HCGs sample $f(E)\sim 30$ per cent, supporting the view that SHKs are richer in early-type galaxies than 
the average Hickson group.

\begin{figure*}
\begin{center}
\includegraphics[width=0.48\textwidth,clip,angle=0]{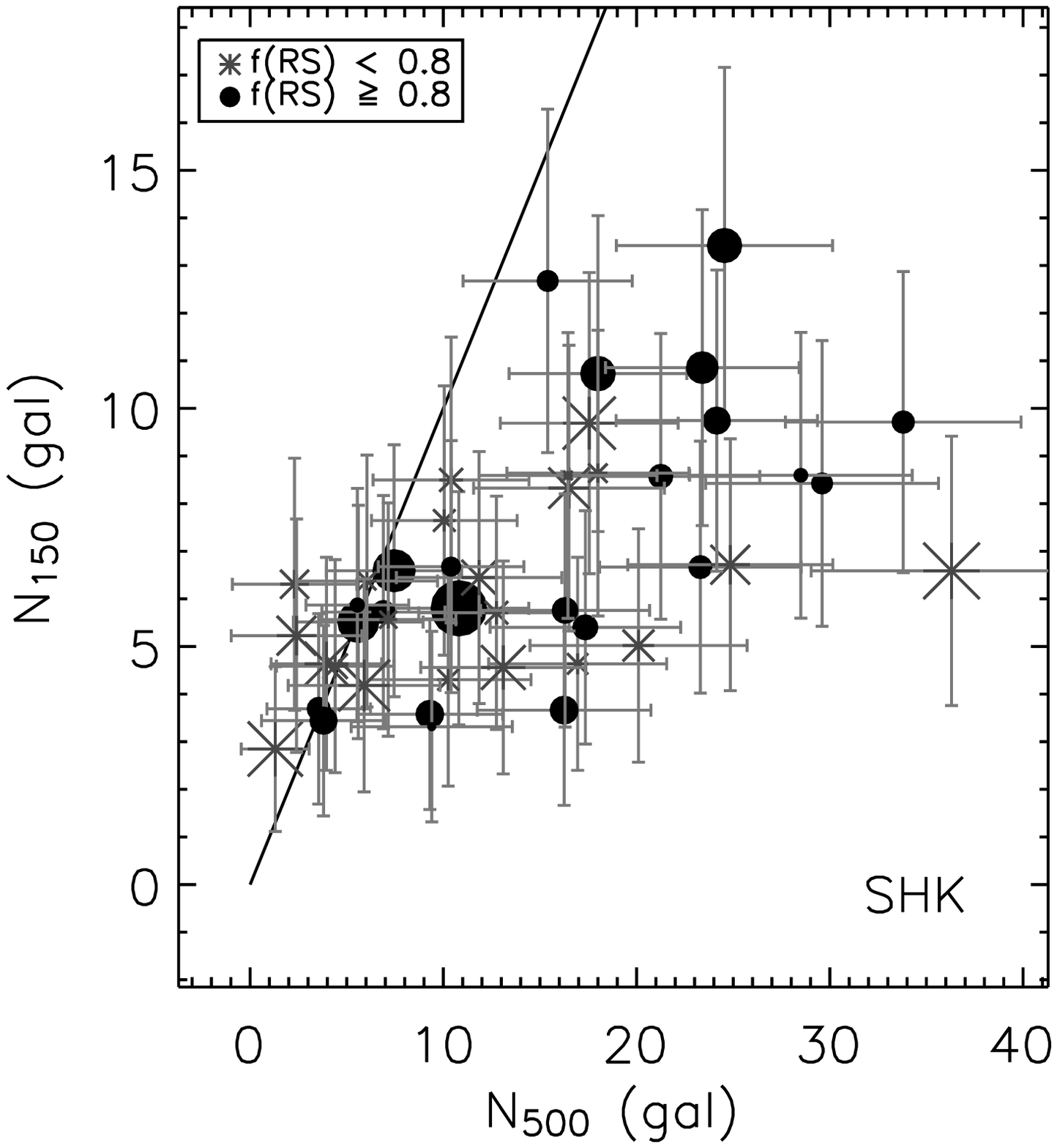} 
\hfill
\includegraphics[width=0.48\textwidth,clip,angle=0]{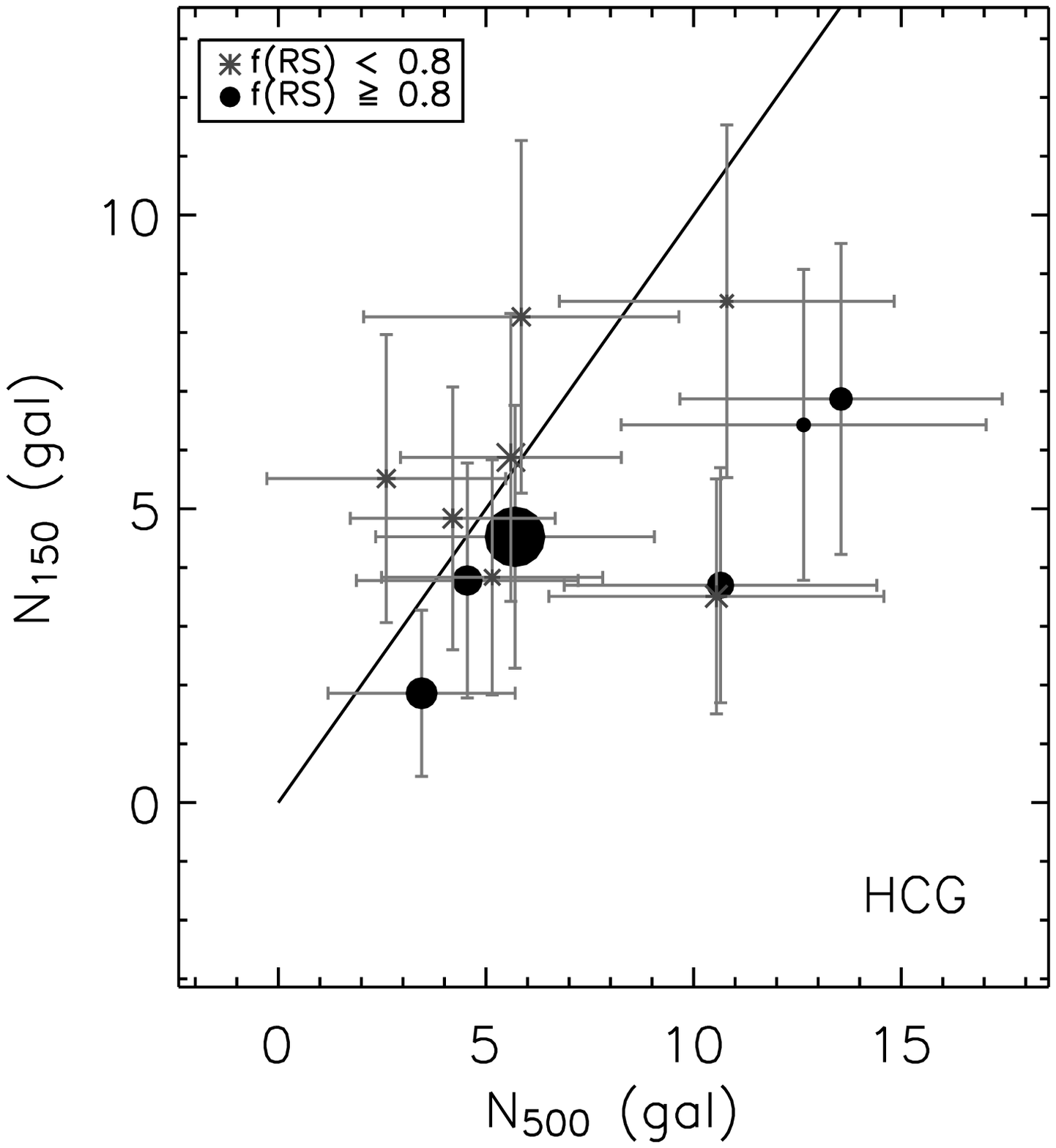} 
\caption{$N_{150}$ richness (see text) estimated within a $ 150 \ {\rm kpc}$ radius versus $N_{500}$ richness within 
$500\ {\rm kpc}$ for the SHKs (left panel) and for the HCGs (left panel). Groups with $f(RS) < 0.8$ and 
$f(RS) \geq 0.8$ are plotted, respectively, with grey asterisks and black dots. The size of the symbols is scaled 
according to the groups mean spectroscopic redshift.}
\label{fig:Fig10}
\end{center}
\end{figure*}

In order to investigate the degree of compactness and isolation of the SHKs, in Fig.~\ref{fig:Fig10} we 
compare the richness estimates measured within $150\ {\rm kpc}$ and $500\ {\rm kpc}$. Despite the 
large uncertainties, our analysis discloses the existence of two classes of groups. The first class is populated 
by objects with $N_{500}\simeq N_{150}$, {\rm i.e.} fairly compact and isolated groups, whose richness does not increase 
going from $150$ to $500\ {\rm kpc}$. The second class is instead populated by more extended and dispersed 
structures for which $N_{500} > N_{150}$; these are on average richer structures, located in the upper right 
part of the left panel in Fig.~\ref{fig:Fig10}.
On the other hand, few HCGs of our sample show the presence of comparable extended structures in their surroundings 
(see right panel in Fig. \ref{fig:Fig10}). 
This may appear to contradict some previous studies on Hickson groups, such as \citet{Rood-1989} or 
\citet{Mamon-1990} who claimed that most HCGs reside within more extended structures. 
More recent studies however \citep{Vennik-1993, Palumbo-1995, deCarvalho-2000}, relying on spectroscopic data-sets, 
found that only part of HCGs are embedded in loose groups. Furthermore we need to point out that our analysis 
method and selection criteria are different from these works, so that a direct comparison is misleading. Here we 
only emphasize the relative difference between our SHK and HCG samples. On the other hand our findings are 
consistent with \cite{Tovmassian-1999a,Tovmassian-2001a,Tovmassian-2001b,Tovmassian-2001c} 
who show that SHKs appear to be embedded into large, loose structures. Their mean value of the 
mass-to-luminosity ratio ($M/L \sim 32\pm 29$ \citep{Tovmassian-2007}, consistent with what found in HCGs 
$M/L \sim 35$ \citep{Hickson-1992}), is though sensibly lower than what usually observed in clusters $M/L \sim 210$ 
\citep{Bahcall-1999}.

\begin{figure*}
\begin{center}
\includegraphics[width=0.48\textwidth,clip,angle=0]{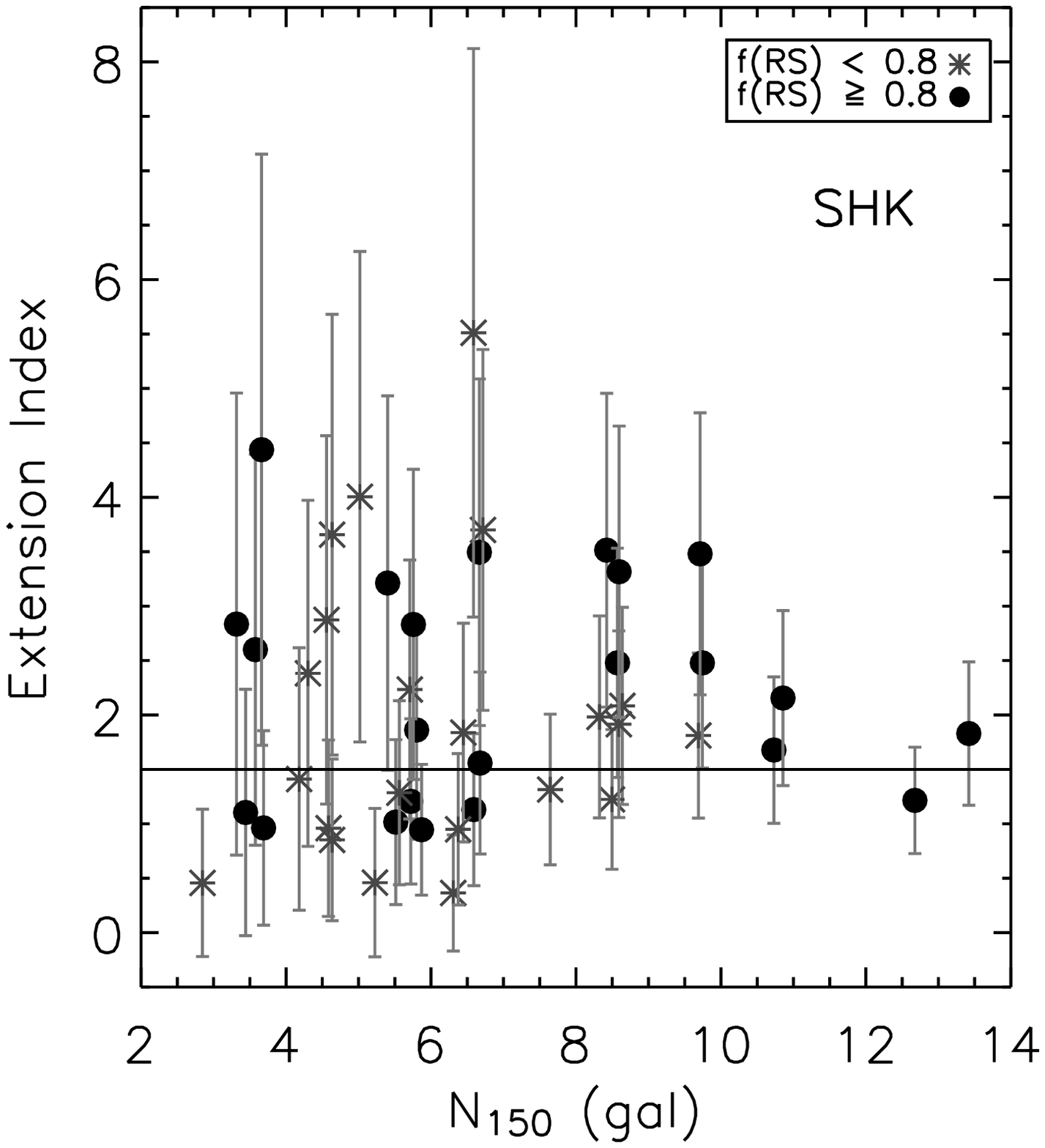} 
\hfill
\includegraphics[width=0.48\textwidth,clip,angle=0]{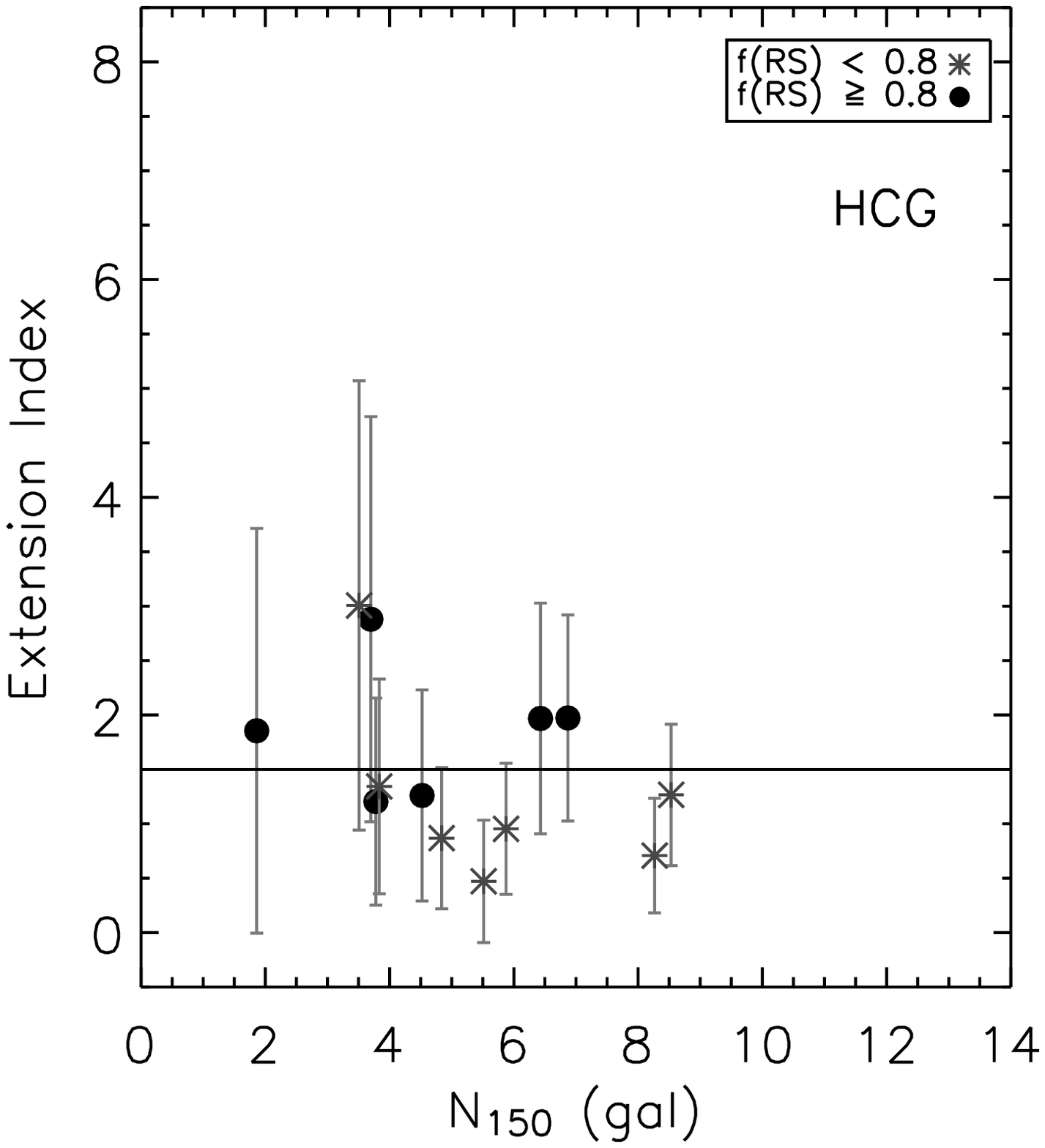} 
\hfill
\caption{Extension index vs $N_{150}$ for SHKs (left panel) and HCGs (right panel). Solid line for
extension index = 1.5 is plotted. Groups with $f(RS) < 0.8$ and $f(RS) \geq 0.8$ are plotted, respectively, with
grey asterisks and black dots.}
\label{fig:Fig11} 
\end{center} 						       
\end{figure*}

We further divided the sample in {\it compact} and {\it extended} groups, depending on their
{\it extension index} defined as {\rm EI}~$=N_{500}/N_{150}$. A trend for the SHKs can be observed with 
increasing richness (see left panel in Fig.~\ref{fig:Fig11}): the majority of rich groups (with 
$N_{150} \geq 7$) is embedded within extended structures (${\rm EI} \geq 1.5$), while poorer structures (with 
$N_{150} < 7$) are a mixture of compact and extended objects (likely [core+halo] configurations). A K-S test 
indicates that the two distributions of {\rm EI} are different with a probability of $94$ per cent. The HCG sample 
instead (Fig.~\ref{fig:Fig11}, right panel) is dominated by concentrated structures, as expected from the strict 
isolation selection criterion used to identify them: less than $15$ per cent of the sample has a concentration 
index larger than 2, to be compared with $>40$ per cent for SHKs. 

\begin{figure}
\begin{center}
\includegraphics[width=0.44\textwidth,angle=0]{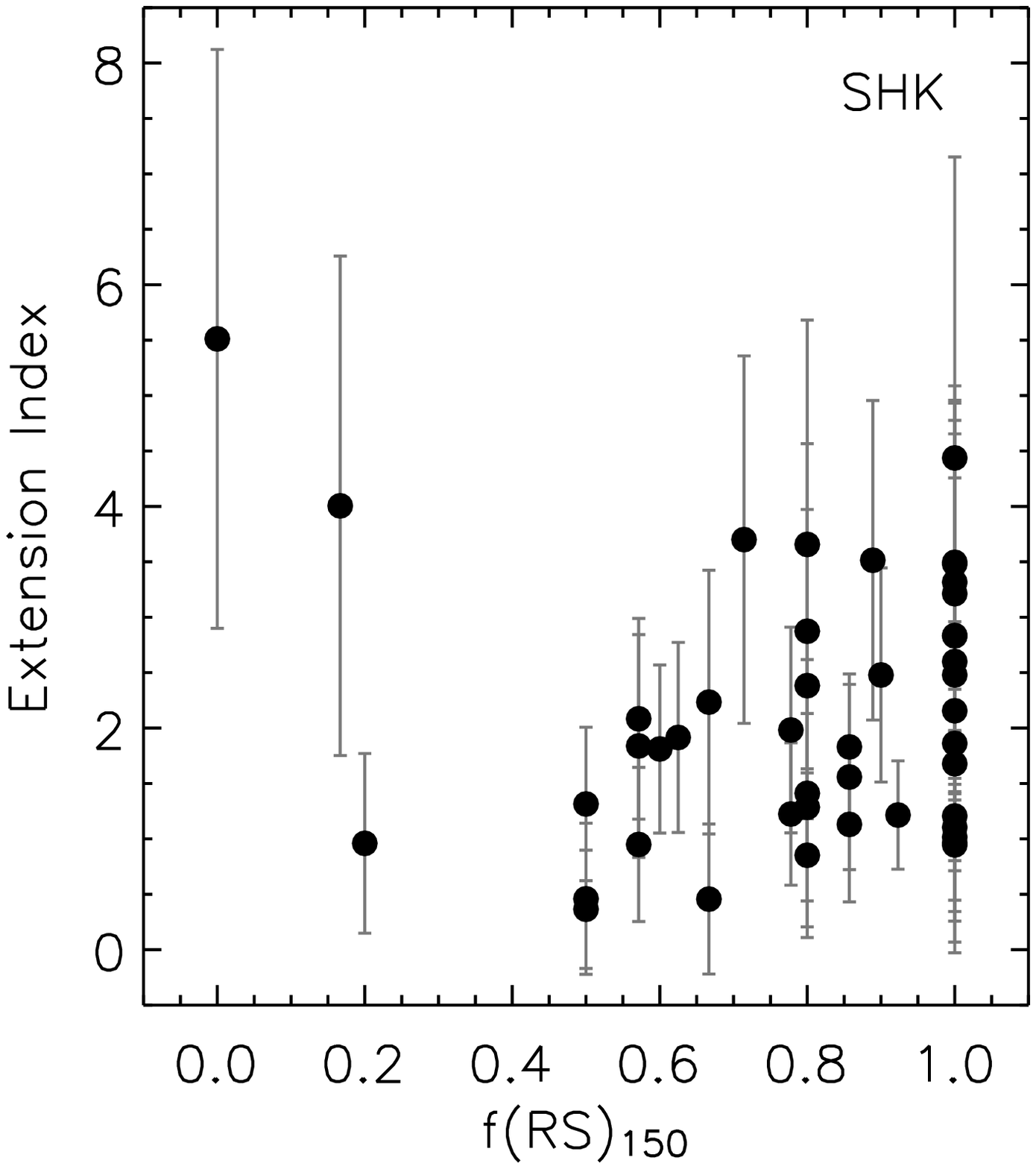}
 \caption{Extension index vs red--sequence galaxies fraction for SHK groups.}
\label{fig:Fig12}
\end{center}  						       
\end{figure}

In Fig.~\ref{fig:Fig10} the size of the symbols is scaled accordingly to the group redshift. No trend with 
redshift is detected. 
Similarly, while we find no statistically significant difference in early-type fraction between more concentrated 
and extended SHK groups, Fig. \ref{fig:Fig12} shows that extended structures are skewed toward higher $f(RS)_{150}$, 
mainly due to the richest systems ($N_{150}\geq 7$) as further discussed in $\S$ \ref{sec:conclusions}. Anyway, we found 
no significant trend between Extension Index and early-type fraction (Spearman's rank test significance value 
of 0.36).

\begin{figure}
\begin{center}
\includegraphics[width=0.44\textwidth,angle=0]{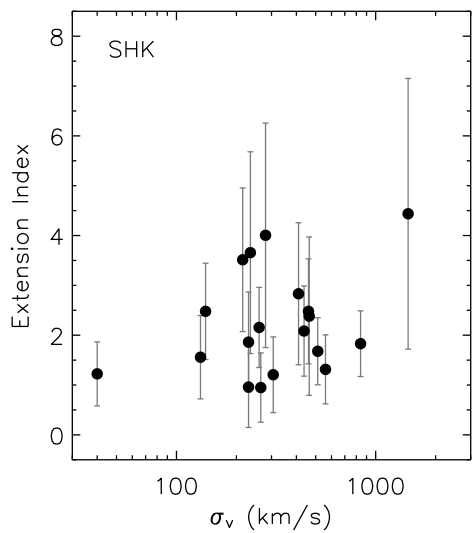} 
\caption{Extension index vs $\sigma_{v}$ for SHK groups. A logarithmic scale on x-axis is used.}
\label{fig:Fig13} 
\end{center} 						       
\end{figure}

To further probe the correlation of the group extension index and of their $f(E)$ on the mass of the system we 
extracted from the literature all available velocity dispersions for our sample of SHKs. Despite the poor 
statistics, due to the very low spectroscopic completeness for these objects, and the fact that no significant 
trend between {\rm EI} and $\sigma_{v}$ is found (Spearman's rank test significance value of 0.49), there is a hint that 
extended groups have on average a higher velocity dispersion ($\overline{\sigma}_{v}=434\pm 93\ {\rm km\ s^{-1}}$) 
(are on average more massive) than isolated groups ($\overline{\sigma}_{v}=280\pm 83\ {\rm km\ s^{-1}}$) suggesting 
that our SHKs sample includes rich extended systems (Fig. \ref{fig:Fig13}).
SHKs structure typologies are summarized in Table~\ref{tab:Table3} together with some of their properties.

\begin{table*}
\begin{tabular}{lccccccc}
\hline
\hline
 {\bf SHK} &  {${\mathbf{f(RS)_{150}}}$}  & {${\mathbf{\Sigma_{500}\ (gal/Mpc^{2})}}$} & 
 {${\mathbf{\Sigma_{150}\ (gal/Mpc^{2})}}$} & {$\mathbf{r^{0}_{1}\ (mag)}$} & {${\mathbf{\Delta_{m}\ (mag)}}$} & 
 {\rm {\bf EI}} & {\bf Structure Typology}\\
  1     &     2   &    3       &     4        &     5	 &    6     &	  7	&	 8		     \\ 								   
\hline
  1     &   0.86  &  31$\pm$7	&   190$\pm$53  &  16.50   &   2.89   &   1.83   &   cluster-core	       \\												
  5     &   1.00  &  7 $\pm$4	&   78 $\pm$35  &  16.38   &   2.48   &   1.01   &   isolated-compact	       \\												
  6     &   1.00  &  30$\pm$7	&   94 $\pm$37  &  15.85   &   2.85   &   3.50   &   core+halo/loose           \\												
  10    &   0.71  &  32$\pm$7	&   95 $\pm$37  &  16.31   &   2.95   &   3.70   &   core+halo/loose           \\												
  11    &   1.00  &  12$\pm$5	&   51 $\pm$28  &  16.67   &   0.54   &   2.60   &   core+halo/loose           \\												
  14    &   0.50  &  13$\pm$5	&   108$\pm$40  &  15.63   &   2.32   &   1.31   &   isolated-compact$^{\ast}$  \\												
  19    &   0.80  &  22$\pm$6	&   66 $\pm$32  &  16.26   &   2.97   &   3.66   &   core+halo/loose           \\												
  22    &   0.20  &  6 $\pm$4	&   65 $\pm$32  &  15.87   &   2.74   &   0.96   &   isolated-compact	       \\												
  31    &   1.00  &  14$\pm$5	&   82 $\pm$35  &  16.36   &   2.24   &   1.86   &   core+halo/loose           \\												
  54    &   1.00  &  22$\pm$6	&   76 $\pm$35  &  15.54   &   2.66   &   3.21   &   core+halo/loose           \\												
  55    &   0.86  &  9 $\pm$4	&   93 $\pm$37  &  16.89   &   2.57   &   1.13   &   isolated-compact	       \\												
  57    &   0.60  &  22$\pm$6	&   137$\pm$45  &  16.48   &   2.68   &   1.81   &   cluster-core	       \\												
  63    &   0.67  &  2 $\pm$2	&   40 $\pm$24  &  17.47   &   1.10   &   0.46   &   isolated-compact	       \\												
  65    &   0.00  &  46$\pm$9	&   93 $\pm$40  &  18.81   &   1.91   &   5.51   &   core+halo/loose           \\												
  74    &   0.17  &  26$\pm$7	&   71 $\pm$35  &  17.54   &   2.69   &   4.00   &   core+halo/loose           \\												
  95    &   1.00  &  4 $\pm$3	&   52 $\pm$28  &  15.18   &   2.50   &   0.96   &   isolated-compact	       \\												
  120   &   0.80  &  13$\pm$5	&   61 $\pm$32  &  16.39   &   2.95   &   2.38   &   core+halo/loose           \\												
  123   &   0.57  &  15$\pm$5	&   91 $\pm$37  &  16.44   &   2.91   &   1.84   &   core+halo/loose           \\												
  128   &   0.80  &  5 $\pm$4	&   66 $\pm$32  &  16.79   &   1.61   &   0.85   &   isolated-compact	       \\												
  154   &   0.89  &  38$\pm$8	&   119$\pm$42  &  15.26   &   2.80   &   3.51   &   cluster-core	       \\												
  181   &   0.90  &  31$\pm$7	&   138$\pm$45  &  15.62   &   2.56   &   2.48   &   cluster-core	       \\												
  186   &   0.67  &  16$\pm$5	&   81 $\pm$35  &  16.06   &   1.15   &   2.23   &   core+halo/loose           \\												
  188   &   1.00  &  9 $\pm$4	&   81 $\pm$35  &  15.04   &   2.09   &   1.21   &   isolated-compact	       \\												
  191   &   1.00  &  23$\pm$6	&   152$\pm$47  &  15.70   &   2.39   &   1.68   &   cluster-core	       \\												
  205   &   1.00  &  21$\pm$6	&   52 $\pm$28  &  16.04   &   0.52   &   4.44   &   core+halo/loose           \\												
  213   &   0.80  &  9 $\pm$4	&   79 $\pm$35  &  14.50   &   2.55   &   1.28   &   isolated-compact	       \\												
  218   &   0.50  &  3 $\pm$4	&   89 $\pm$37  &  16.28   &   2.73   &   0.36   &   isolated-compact	       \\												
  223   &   1.00  &  27$\pm$6	&   121$\pm$42  &  15.44   &   2.23   &   2.48   &   cluster-core	       \\												
  231   &   0.80  &  17$\pm$5	&   64 $\pm$32  &  16.74   &   2.64   &   2.87   &   core+halo/loose           \\												
  237   &   1.00  &  12$\pm$5	&   47 $\pm$28  &  14.34   &   2.84   &   2.83   &   core+halo/loose           \\												
  245   &   0.57  &  23$\pm$6	&   122$\pm$42  &  14.94   &   1.96   &   2.08   &   cluster-core	       \\												
  251   &   0.57  &  8 $\pm$5	&   90 $\pm$37  &  16.00   &   2.74   &   0.95   &   isolated-compact	       \\												
  253   &   0.92  &  20$\pm$6	&   179$\pm$51  &  15.53   &   1.95   &   1.21   &   isolated-compact$^{\ast}$  \\												
  254   &   0.80  &  7 $\pm$5	&   59 $\pm$32  &  17.54   &   2.71   &   1.41   &   isolated-compact	       \\												
  344   &   0.78  &  13$\pm$5	&   120$\pm$42  &  15.84   &   2.78   &   1.22   &   isolated-compact$^{\ast}$  \\												
  346   &   0.78  &  21$\pm$6	&   118$\pm$42  &  16.36   &   2.64   &   1.98   &   cluster-core	       \\												
  348   &   1.00  &  21$\pm$6	&   81 $\pm$35  &  15.33   &   2.85   &   2.83   &   core+halo/loose           \\												
  351   &   0.62  &  21$\pm$6	&   121$\pm$42  &  13.60   &   2.76   &   1.91   &   cluster-core	       \\												
  352   &   1.00  &  36$\pm$7	&   122$\pm$42  &  14.11   &   2.82   &   3.32   &   cluster-core	       \\												
  355   &   1.00  &  5 $\pm$4	&   49 $\pm$28  &  16.49   &   1.97   &   1.10   &   isolated-compact	       \\												
  357   &   1.00  &  43$\pm$8	&   137$\pm$45  &  14.70   &   2.86   &   3.48   &   cluster-core	       \\												
  358   &   1.00  &  7 $\pm$3	&   83 $\pm$35  &  13.36   &   2.93   &   0.94   &   isolated-compact	       \\												
  360   &   1.00  &  30$\pm$6	&   154$\pm$47  &  15.48   &   2.81   &   2.15   &   cluster-core	       \\												
  371   &   0.50  &  3 $\pm$4	&   74 $\pm$35  &  16.69   &   2.66   &   0.46   &   isolated-compact	       \\												
  376   &   0.86  &  13$\pm$5	&   94 $\pm$37  &  14.73   &   2.46   &   1.56   &   core+halo/loose           \\												
\hline														
\end{tabular}								  					
\caption{													
Column 1: SHK identification number;											
column 2: Red Sequence galaxies fraction within the inner region;							   										 
column 3: number spatial density inside a radial distance of $500\ {\rm kpc}$, determined using $N_{500}$; 		   			
column 4: number spatial density inside a radial distance of $150\ {\rm kpc}$, determined using $N_{150}$;  
column 5: group's brightest galaxy as defined in \S \ref{sec:themethod}; 
column 6: magnitude gap between the brightest and the faintest galaxies inside a radial distance of $150\ {\rm kpc}$;
column 7: group's extension index as defined in \S \ref{subsec:global}; 						    
column 8: Structure Typology. 
The asterisks highlight the richest compact structures ($N_{150}\geq7$).}
\label{tab:Table3}	  
\end{table*}

\subsection{Shakhbazyan Groups in X-Rays}
\label{subsec:SHK-X-Ray}
The X-ray emission due to hot ionized gas in galaxy groups or clusters is an indication of the 
presence of a common potential well, and thus of the physical reality of the structure.
Unfortunately very few pointed observations of SHK groups in the X-ray band are available in the archives. 
Thus, in order to investigate the X-ray luminosity of these systems, we combined pointed and Rosat All Sky Survey 
({\it RASS}) data. Excluding all groups which fall too far from the nominal aim-point of the satellite (and thus are affected 
by large uncertainties), we are left with 4 groups with pointed observations: SHK 37 and 154 observed by {\it Chandra} and 
{\it XMM}, and SHK 202 and 233 \citep{Takahashi-2000,Takahashi-2001} observed with {\it ASCA}, while 8 groups (1,154,202,346,352,355,357,360) are individually 
detected in the {\it RASS} observations. In cases where both pointed and {\it RASS} data were available we verified the consistencies of 
the measured fluxes, but decided to use the latter for homogeneity.

For all remaining SHK groups without individual X-ray detection, we stacked the {\it RASS} X-ray photons, obtaining a 
significant detection at the $5\sigma$ level within 5', {\rm i.e.} $\sim 500$ {\rm kpc} at the average redshift of 0.09. 
The use of a more physical extraction aperture would be preferable, but due to the problems discussed in 
Sec. \ref{sec:themethod} that is hardly feasible with the available data. However note that while the extraction 
radius was chosen primarily to allow an easy comparison with the literature (see below), the adopted aperture is 
close to the average virial radius for a group with velocity dispersion of ~250 ${\rm km\ s^{-1}}$ and our 
redshift range.

In order to verify the likelihood of a fake detection we further performed 100 additional stackings choosing 
random fields within 1 and 2 deg from the group centroids. We find that the likelihood of a chance detection is 
$<3$ per cent. In Table \ref{tab:Table4} we report the average luminosity of the whole sample, as well as those of 
the sub-samples selected according to $f(E)$, concentration and velocity dispersion $\sigma_{v}$ (when available).

\begin{figure}
\includegraphics[width=0.5\textwidth,angle=0]{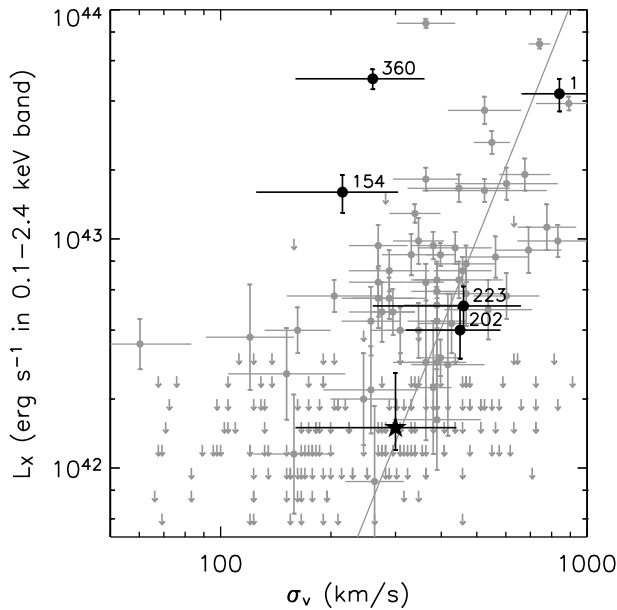} 
\caption{$L_{X}$ vs $\sigma_{v}$ distribution of SHK groups. Full black dots represent individual detections, while 
the star shows the luminosity of the stacked sub-sample of undetected SHK groups and the range spanned by 
different sub-samples in Tab.\ref{tab:Table4}. Grey dots and arrows show the group sample of Mahdavi et al. (2000) 
while the solid line is the extrapolation of the relation measured for rich clusters.}
\label{fig:Fig14}  						       
\end{figure}

Our results are compared with the $L_X$ vs $\sigma_{v}$ plot of \cite{Mahdavi-2000}, based on {\it RASS} data, in Figure 
\ref{fig:Fig14}. We can see that X-ray luminosities for SHK 1, 202 and 223 are consistent with the expectations 
for groups of comparable $\sigma_{v}$, even though on the faint side. SHK 154 and 360 are brighter than expected, 
suggesting that their velocity dispersions were underestimated since both of them are part of rich clusters.
The remaining undetected groups also have an average luminosity consistent with the expectations, assuming a 
median $\sigma_{v}\simeq 300\ {\rm km\ s^{-1}}$, with a dependence on $f(E)$, concentration and $\sigma_{v}$ in 
agreement with 
the expected trends\footnote{Note that our estimates are upper limits to the actual emission of diffuse intragroup 
gas, due to the impossibility of disentangling the overall emission from the one of individual galaxies and/or 
AGNs. However given the few ($<5$) bright early-type galaxies within the extraction radius, we estimate 
that the contribution of individual sources is on average less than a few $10^{41}$ erg s$^{-1}$.} 
(see Table~\ref{tab:Table4}).
While the adoption of a fixed radius for all sources may increase the scatter on the determination of the 
X-ray properties of SHK groups (but not in the comparison with Mahdavi and collaborators), the uncertainties due 
to the poor statistics, the gas temperature,  the AGN contribution {\rm etc.}, represent the dominant source of 
uncertainty.

\begin{table}
\begin{center}
\begin{tabular}{lc}
\hline
{Sub-sample} & {$L_X$($10^{42}$ erg s$^{-1}$)}\\
\hline
All undetected & $1.5\pm 0.3$\\
high $f(E)$ & $2.3\pm 0.5$\\
low $f(E)$ & $1.2\pm 0.4$\\
{\rm EI} $\geq1.5$ & $1.8\pm 0.5$\\
{\rm EI} $<1.5$ & $1.4\pm 0.4$\\
$\sigma_{v}>300\ {\rm km\ s^{-1}}$ & $2.6\pm 0.8$\\
$\sigma_{v}<300\ {\rm km\ s^{-1}}$ & $1.3\pm 0.6$\\
\hline
\end{tabular}
\caption{Average stacked X-ray luminosities in the 0.1-2.4 {\rm keV} band within 500 {\rm kpc} from {\it RASS} 
data for groups which are individually undetected in X-rays.}
\label{tab:Table4}
\end{center}
\end{table}

\section{Discussion and Conclusions}
\label{sec:conclusions}
Shakhbazyan groups of galaxies, in spite of having been originally selected as ``compact groups of compact
galaxies'', due to the less restrictive selection criteria adopted by their discoverers, have been shown to
sample a large range of spatial density. In an ongoing effort to study the low density regime of cosmic
structures in the local Universe we have performed a systematic study of all the SHKs covered by the Sloan
Digital Sky Survey DR5. Unfortunately only a minority of SHKs have measured spectroscopic redshifts and often
only for one or two galaxies in each group. However, using a series of diagnostics in 2D and in photometric 
redshift space, derived with neural networks tools, we found that $\geq 78$ per cent of 
them are confirmed structures with richness ranging from $3$ to $13$ galaxies.

This result is intriguing in light of the fact that SHKs are missing in other catalogues of galaxy groups. Only 
2 of the SHK groups inside the SDSS sky coverage are in common with the Hickson sample. 
Several justifications could be provided about why Hickson missed most of the compact SHK groups:
i) Hickson looked for much more compact structures than SHKs are, with galaxies' relative distances lower than 
the characteristic diameter of a member galaxy while SHKs are selected with a less restrictive criterion, hence
they may or not comprehend HCG-like structures. This also reflects on the high threshold imposed by Hickson's 
surface brightness criterion, which is often not satisfied by SHKs;
ii) The galaxies compactness selection criterion used to identify SHKs renders them easily
contaminated with stars misclassified as galaxies. The opposite can naturally also easily happen;
some SHKs could hence have been misclassified as star clusters, especially the high redshift ones;
iii) The isolation criterion used by Hickson to identify his groups is often
unsatisfied by SHKs, since, as we have found in our analysis, even the most compact structures are embedded in 
looser ones. Furthermore, a recent analysis of galaxies in cosmological simulations indicates that the HCG 
sample is highly incomplete \citep{Diaz-2008}.\\ 
\noindent It is worth pointing out that our definition of compactness is different 
from the one used by Hickson, and they may coincide only for the most compact SHKs.

Notwithstanding the difficulty in reproducing Hickson's selection criteria accurately (see {\rm e.g.}
\citealp{Iovino-1997}), we have applied them to our SHKs sample using the same magnitude limit that 
characterized the original POSS plates. We found that they are satisfied only by 9 groups (SHK 8, 19, 95, 128, 
237, 248, 355, 358, 371) out of 58, 7 of which are identified as likely real structures by our procedure. Five of 
these (SHK 95, 128, 355, 358, 371) are classified as compact structures (${\rm EI}<1.5$), while the other two (SHK 19, 
237) are likely [core+halo] configurations (${\rm EI}\geq1.5$ and $N_{150}<7$)).\\  

The overlap with other published studies of galaxy groups is also extremely poor 
({\rm e.g.} SDSS: \citealp{Tago-2008, Berlind-2006}; UZC: \citealp{Focardi-2002}; also see \citealp{Lee-2004}). This is most 
probably due to the fact that they are not as isolated and thus are entirely missed by detection algorithms 
looking for isolated and compact structures. Also, so far, almost all studies of groups in the SDSS have been based on 
the spectroscopic data-set, which is highly incomplete for environments as dense as the Shakhbazyan groups.
In fact we find that only a minority ($<10$ per cent) of candidate group members belongs to the SDSS 
spectroscopic sample. However, see \citet{McConnachie-2008b} for a study of compact and isolated HCG-like galaxy 
groups based upon the SDSS photometric data-set. 

Using photometric ({\rm e.g.} \citealp{Strateva-2001}) and CMR based criteria, we studied the morphological content of 
SHKs and found that more than $90$ per cent  and $75$ per cent of SHKs have indeed early-type fractions 
$>0.6$, using the {\it u-r} or Red Sequence methods respectively. 
The overabundance of early-type and Red Sequence galaxies is however confined to the group core since when we 
consider the morphological content of the local environment ($150<r<1000\ {\rm kpc}$) and background 
($2<r<3\ {\rm Mpc}$) we observe that red galaxies decreases outward. 
Following the same approach, we also estimated early-type and Red Sequence fractions for a sub-sample composed of 
all HCGs with $z_{{\rm spec}} \geq 0.03$, located within the SDSS. When compared with our sub-sample of HCGs, we found 
that SHKs have somewhat higher values of $f(E)$, especially considering that our HCG sub-sample is biased toward 
high early-type fractions.
The large fraction of early-type gas-poor galaxies found in SHKs, could partly explain the fact that enhanced 
FIR emission was detected in only a small fraction ($\sim7$ per cent) of them~\citep{Tovmassian-1998}, in contrast to the  
more pronounced excess found in HCGs ($>60$ per cent)~\citep{Allam-1996}. 

Our analysis discloses the existence of two classes of SHKs, one constituted by compact and isolated groups
(${\rm EI}=N_{500}/N_{150}<1.5$) and the other by dispersed and more extended structures (${\rm EI}\geq 1.5$). A trend 
for SHKs can be observed for the extension index and the inner early-type fraction with increasing richness: 
rich SHKs (with $N_{150}\geq 7$) are embedded within extended structures (${\rm EI}\geq 1.5$), while poorer ones 
(with $N_{150}< 7$) are a mixture of isolated and more extended objects. The richest groups tend to be dominated by
early-type galaxies ($f(E)_{150}\geq 0.8$). On the other hand, we find that the analysed HCGs are composed only of 
poor ($N_{150}< 7$), concentrated (${\rm EI}<1.5$) structures: less than $15$ per cent of the sample has an extension index 
larger than 2, to be compared with $>40$ per cent for SHKs. We point out that this trend is not expected just based 
on the selection criteria since the isolation constrain tends to increase the {\rm EI} for groups with radius 
$R_{\rm G}<50\ {\rm kpc}$\footnote{Half of the angular diameter $\theta_{\rm G}$ of HCGs as found by Hickson's selection 
criteria, measured in {\rm kpc}.} ($3 R_{\rm G}<150\ {\rm kpc}$), i.e. for half of the overall HCG sample and 40 per cent of ours. 
Furthermore the majority of SHKs satisfying the HCG criteria has ${\rm EI}<1.5$ (5 out of 7). This implies that HCG criteria 
effectively select compact and truly isolated groups which are deficient in galaxies also outside the avoidance region.
Isolated SHKs (${\rm EI}<1.5$) dwell in less dense regions such as the outer parts of clusters or the field, possibly 
sharing several properties of the Hickson sample.
Dispersed SHKs (${\rm EI} \geq 1.5$) are a mixture of different classes. Those having $N_{150}< 7$ are probably 
[core+halo] configurations or condensations within larger structures.
Those with richness $N_{150}\geq 7$ are extended structures whose central content of red galaxies is high
because we are considering the central cores of galaxy clusters. In fact all SHKs which we were able to associate 
with known Abell or Zwicky clusters belong to the latter group ({\rm e.g.}, SHK 360 is the central part of 
Abell 2113 cluster, SHK 191 is also classified as Abell 1097). According to what we have found for SHKs and 
HCGs, it seems that there are almost no [core+halo] configurations among our HCGs comparable to the class of SHK 
groups with large {\rm EI}.
 
Our analysis thus suggests that, contrary to what previously thought (see for instance \citealp{Tiersch-2002}), 
SHKs are an extremely heterogeneous class of objects which includes both cores of rich, extended structures and 
compact, isolated groups. The differences with HCGs are mainly due to the contamination of the sample by cores of
galaxy clusters which have no counterpart in the HCG sample. This is also likely the cause of the SHKs bias toward 
larger early-type fractions. This scenario is further supported by the fact that extended SHKs have in average 
higher velocity dispersion ($\overline{\sigma}_{v}=434\pm 93\ {\rm km\ s^{-1}}$) than compact ones 
($\overline{\sigma}_{v}=280\pm 83\ {\rm km\ s^{-1}}$).
Furthermore, while the velocity dispersions of SHKs (where available) appear to be higher ($\gtrsim 300\ {\rm km\ s^{-1}}$) 
and crossing times smaller ($t_c\sim 90$ {\rm Myr}), than those of most compact groups, neglecting the presence of an 
extended halo inevitably leads to underestimate the crossing times. Our results are also in agreement with the 
simulations performed by \citet{McConnachie-2008a}, according to which compact groups, although selected to be 
compact and isolated, are often the cores of larger groups.

We didn't find any correlation between the environment in which SHKs reside and their redshift. This suggests
that the evolutionary state of these groups doesn't depend on their cosmological age (on the redshift)
but only on the density of the zone in which they are born. However we must consider that the redshift range 
covered by our sample ($0.02\lesssim z \lesssim 0.3$) may not be enough to detect evolutionary trends.

In addition, notwithstanding the fact SHKs original selection criteria may potentially produce
a heterogeneous collection of loose nearby groups and denser distant ones, the overall properties and in 
particular the compact/sparse nature of SHKs in our sample do not appear to depend on their distance. In fact, we 
found no significant redshift dependence of the magnitude difference between the faintest and the brightest galaxy 
in each group ($\Delta_{m}=m_{\rm faint}-m_{\rm bright}$), of the Extension Index ({\rm EI}), of the surface number density or 
the X-ray luminosity. 
Neither did we find any correlation between $\Delta_{m}$ and any of the other parameters.
Note however that our selection criteria to define both the optimal SHK sample and the member galaxies, may 
contribute to reduce any bias present in the original SHK catalogue.
    
Finally, X-ray luminosities for SHKs detected in X-rays are consistent with the expectations for the 
$L_{X}-\sigma_{v}$ relation. The remaining undetected groups also have an average luminosity consistent with the 
expectations, assuming a median $\sigma_v\simeq 300\ {\rm km\ s^{-1}}$.

SHKs have not been intensely investigated and still little is known about their properties and state of evolution.
Available data for SHKs are of poor quality: the few detailed works available in literature are devoted to the 
rich end of the catalogue, and 3/4 of the groups with robust velocity dispersion belong to the extended
sub-sample. Most of the claims (relatively high velocity dispersions, short crossing times, {\rm etc.}) have hence to be
revisited, since they are biased towards richer structures, which are possibly also the more massive ones, 
reside in peculiar environments and do not share the properties of the whole sample so that they cannot be easily
compared with HCGs. 

The bias introduced by Shakhbazyan in looking for dense systems made by very red galaxies with
the aim to find a homogeneous class of structures, resulted in the fortuitous identification of a more complicated 
situation constituted of different kinds of systems, the poorest of them are probably going through different 
evolutionary state. The high content of early-type galaxies in these structures is due to an 
environmental--dynamical effect that is simply the byproduct of the Morphology--Density relation.   

To properly characterize the properties of these groups and their evolutionary path, it is essential to obtain 
more accurate redshift determination for those objects for which no spectroscopic data exist in literature 
($80$ per cent of the complete SHKs catalogue) and uniformly sample the different sub-populations. This will allow to 
confirm candidate group members and hence to improve the richness and dynamical properties estimates of the groups.
Group masses and {\it M/L} ratios will eventually allow to put limits on the dark matter content, dynamical time--scales 
and evolutionary state of these structures. Additional X-ray observations would provide an additional proof to the physical 
reality of SHKs and would allow to estimate the total mass and the baryonic fraction of these systems.

\subsection*{Acknowledgments}
We thank the referee, Dr. Mamon, for his valuable comments which allowed to significantly improve the quality of
this paper. DC also expresses his gratitude to Chris A. Collins for the useful discussions and encouragements he 
has provided.\\

\noindent This work was partly funded by the Italian Ministry of Foreign Affairs (MAI) through a bylateral Italy-USA agreement.\\

\noindent The authors express their strongest disagreement with the unsupportive attitude
shown over the last twenty years, by the various Italian Governments,
towards scientific research and public universities, and their concern about 
the latest reforms and funding cuts which are undermining the career of young 
researchers.\\

\noindent Funding for the Sloan Digital Sky Survey (SDSS) and SDSS-II has been
provided by the Alfred P. Sloan Foundation, the Participating
Institutions, the National Science Foundation, the U.S. Department of
Energy, the National Aeronautics and Space Administration, the Japanese
Monbukagakusho, the Max Planck Society and the Higher Education
Funding Council for England. The SDSS Web site is http://www.sdss.org/.

\bibliographystyle{mn2e}

\bibliography{Reference}

\end{document}